\documentclass[12pt]{article}

\usepackage[pdftex]{graphicx} 
\usepackage{jheppub}
\usepackage{amsfonts,amsmath} 
\usepackage{mathrsfs} 
\usepackage{setspace}
\usepackage[utf8]{inputenc} 
\usepackage{comment} 

\makeatletter
\def\@fpheader{\relax}
\makeatother

\newcommand{\be}{\begin{equation}} \newcommand{\ee}{\end{equation}}
 
\newcommand{\half}{{\textstyle \frac{1}{2}}}

 \newcommand{\goesto}{\rightarrow}
 \renewcommand{\Re}{\,{\rm Re}\,}

\DeclareMathOperator{\Tr}{Tr}

\newcommand{\Dp}{\partial_{z}}
\newcommand{\Dt}{\partial_{t}}
\newcommand{\fosum}[2][3]{\sum_{i=1}^{#1}\frac{#2}{z-z_i}}

\newcommand{\Dfrac}[2]{\frac{d#1}{d#2}}

\title{Isomonodromy, Painlevé Transcendents and Scattering off of Black
Holes}

\author{Fábio Novaes and}
\author{Bruno Carneiro da Cunha}
\emailAdd{fabiomnsantos@df.ufpe.br}
\emailAdd{bcunha@df.ufpe.br}
\affiliation{Departamento de Física, Universidade Federal de Pernambuco,
50670-901, Recife, Pernambuco, Brazil} 

\abstract{We apply the method of isomonodromy to study the scattering of
a generic Kerr-NUT-(A)dS black hole. For generic values of the charges,
the problem is related to the connection problem of the Painlevé VI
transcendent. We review a few facts about Painlevé VI, Garnier
systems and the Hamiltonian structure of flat connections in the Riemann
sphere. We then outline a method for computing the scattering
amplitudes based on Hamilton-Jacobi structure of Painlevé, and discuss
the implications of the generic result to black hole complementarity.}

\preprint{\today}

\begin{document}

\maketitle

\section{Introduction}

The scattering of quantum fields around a classical gravitational
background has had an important impact in the study of quantum fields in
curved spaces \cite{BirrellDavies}. The classical result for
Schwarzschild black holes has helped in the development of the Hawking
mechanism, as well as fundamental questions of unitarity of evolution and
the exact character of the event horizon as a special place in a black
hole background. From the classical perspective, one can also study
stability of solutions by these means, treating the metric
perturbations as a spin-2 field. 

With the advent of the gauge-gravity correspondence, black holes became
the prototype of thermal systems, so the problem of scattering had other
applications such as the calculation of normal modes and transport
properties for the dual theory \cite{Son:2007vk,Berti:2009kk}. On a more
mathematical perspective, the black hole scattering is linked to the
monodromy of a Fuchsian equation,
\cite{Motl:2003cd,Castro2013,Castro2013b}, a fact which drew some
attention of late because of its relation to conformal field theory and
Liouville field theory \cite{Alday:2009aq,Alba:2010qc}. A Fuchsian
differential equation is one whose solutions diverge polynomially at
singular points. The ocurrence of Fuchsian equations in mathematical
physics is pervasive. The mathematical structure behind them is rich
enough that one can know enough about their solutions in order to define
new functions which have direct physical application.

This article deals with one such structure, the problem of isomonodromic
deformations, and its relation the problem of scattering of black holes.
In mathematics, the problem was initially studied with the application to
the Riemann-Hilbert problem. The latter consists of finding a Fuchsian
ordinary differential equation (ODE) with prescribed monodromy data. Soon
Poincaré noted that the amount of independent parameters in a Fuchsian
ODE with regular singular points was not sufficient to parametrize a
generic monodromy group, and then instead one began considering a generic
linear system of first order differential equations. Schlesinger
\cite{Schlesinger:1912} found a system of differential equations that
enabled one to change the parameters with respect to the positions of the
singular points in order to keep the monodromy data fixed. These
equations bear his name, and details of the construction can be found in
\cite{Iwasaki:1991}.

The construction above has a direct relation to the problem of
scattering. Generically, a second order linear differential equation will
have two linearly independent solutions. If the differential equation is
Fuchsian, then the solutions near a singular point will be asymptotically
of the form $(z-z_i)^{\rho^\pm_i}$, with the exponents $\rho^\pm_i$
solutions of the indicial equation near $z_i$. These
solutions can be continued via Frobenius construction up to another
singular point $z_j$, where there will be another pair of asymptotic
solutions and exponents. As a second order linear ODE has two linearly
independent solutions, in general the first set of asymptotic solutions
will be a linear combination of the second set. The matrices that relates
any such two pair of solutions is called the monodromy data.  In a
scattering process, if one has, say, a problem of a ``purely
ingoing'' wave in the vicinity of a black hole horizon, then one can use
the monodromy data to relate it to a combination of ``ingoing'' and
``outgoing'' waves at infinity, and from those coefficients one can
compute the scattering amplitudes.

The inverse scattering method has had a very close relationship with
integrable structures. This article tries to take advantage of it in two
ways: first, one should ask which assumptions one has to impose on the
spacetime in order to separate the wave equation and reduce it to a
system of Fuchsian equations. Second, one can ask if (and how) the
isomonodromic flow helps in obtaining the monodromy data for those
Fuchsian equations. As it turns out, the assumptions are exactly that the
spacetime has a principal conformal Killing-Yano tensor, which points to a
twistorial structure. Solutions of Einstein equations with a cosmological
constant with a Killing-Yano tensor are completely determined by their
global charges: its mass, angular momentum, and NUT charge. For the second
point, we stabilish that the monodromy data depends on a modified
$\tau$-function associated to the isomonodromy flow. Incidentally, the
twistorial structure of the isomonodromic flow is more sophisticated and
already studied in \cite{Mason:2000aa}.

The paper is organized as follows. In section two, we perform the
reduction of the Klein-Gordon equation into a pair of Fuchsian equations
with five regular singular points, if the cosmological constant is not
zero. In the case of conformally coupled scalars, the number of singular
points is reduced to four. In section three, we introduce the Schlesinger
and Garnier systems which appear in the isomonodromy applications, and
related the isomonodromic flow to the Painlevé VI system. In section four,
we borrow from the theory of flat non-abelian connections to introduce a
symplectic structure in the space of monodromy parameters. In section
five, we consider the canonical transformation between the Painlevé VI
system and the monodromy parameters. In section six, we discuss the
implications for the scattering of eternal black holes, specifically
between different asymptotic regions, and close with a summary of
results. Some technical results about the monodromy of the
hypergeometric equation and the asymptotics of the Painlevé
system are presented in the appendix.

\section{Killing-Yano and Separability}

\label{sec:hidd-symm-kerr}

A 2-form $h_{ab}$ is a principal conformal Killing-Yano tensor (PCKY)
if it is a closed and non-degenerate conformal Killing-Yano tensor
(CKY), satisfying $\nabla_{(a}h_{b)c}=g_{ab}\eta_c-g_{c(a}\eta_{b)}$ for
$\eta_a=\nabla^bh_{ab}/(D-1)$. Consider a spacetime $(M,g)$ with
$D=2n+\varepsilon$ dimensions allowing a PCKY, where
$\varepsilon=0,1$. The existence of such structure implies a tower of
$n-1$ Killing-Yano tensors, which implies $n$ Killing tensors if we
include the metric tensor. Those killing tensors can then be used to
construct $n+\varepsilon$ commuting killing vectors. Thus a spacetime
with a PCKY has $D=2n+\varepsilon$ conserved quantities. This is
sufficient for integration of the geodesic equation \cite{Wald:1984},
but it is also enough for complete separability of Klein-Gordon, Dirac
and gravitational perturbation equations
\cite{Frolov2007a,Frolov2009a}. 

Following \cite{Frolov2007a}, we can choose canonical coordinates
$\{\psi_i,x_\mu\}$, where $ \psi_0$ is the time coordinate, $\psi_k,\;
k=1,\dots,n-1+\epsilon$, are azimuthal coordinates -- Killing vector
affine parameters, and $x_\mu,\;\mu=1,\dots,n$ stand for radial and
latitude coordinates. In such coordinates the generic metric of $(M,g)$
which allows for a PCKY can be written as 
\begin{equation}
  \label{eq:metricPCKY}
  ds^2 = \sum_{\mu = 1}^{n}
  \left[
    \frac{dx_\mu^2}{Q_\mu} + Q_\mu
    \left(
      \sum_{k = 0}^{n-1}A_\mu^{(k)}d\psi_k
    \right)^2 - \frac{\epsilon c}{A^{(n)}}
    \left(
      \sum_{k = 0}^{n-1}A^{(k)}d\psi_k
    \right)^2
  \right]
\end{equation}
where
\begin{gather}
  \label{eq:metricfunctions}
  Q_\mu = \frac{X_\mu}{U_\mu},\quad A_\mu^{(j)} =
  \sum_{\substack{\nu_1 < \dots < \nu_j \\ \nu_i \neq \mu}} x_{\nu_1}^2
   \dots x_{\nu_j}^2,\quad A^{(j)} =
  \sum_{\substack{\nu_1 < \dots < \nu_j}} x_{\nu_1}^2 \dots
  x_{\nu_j}^2,\\
  U_\mu = \prod_{\nu \neq \mu}(x_\nu^2 - x_\mu^2),\quad X_\mu =
  \sum_{k=\epsilon}^{n}c_kx_\mu^{2k} - 2b_\mu x_\mu^{1-\epsilon}+
  \frac{\epsilon c}{x_\mu^2}.
\end{gather}
The polynomial $X_\mu$ is obtained by substituting the metric
\eqref{eq:metricPCKY} into the $D$-dimensional Einstein
equations. The metric with proper signature is recovered when we set
$r=-ix_n$ and the mass parameter $M = (-i)^{1+\epsilon}b_n$.

One of the most interesting properties of the Kerr-NUT-(A)dS metric is
separability. Consider the massive Klein-Gordon equation
 \begin{equation}
   \label{eq:6}
   (\Box - m^2)\Phi = 0,
 \end{equation}
its solution can be decomposed as
\begin{equation}
  \label{eq:12}
  \Phi = \prod_{\mu=1}^n
  R_\mu(x_\mu)\prod_{k=0}^{n+\epsilon-1}e^{i\Psi_k \psi_k}  
\end{equation}
and substitution in \eqref{eq:6} gives
\begin{equation}
  \label{eq:13}
  (X_\mu R_\mu^{\prime})^{\prime} +
    \epsilon\frac{X_\mu}{x_\mu}R_\mu^{\prime} +
    \left(
      V_\mu - \frac{W_\mu^2}{X_\mu}
    \right)R_\mu = 0,
\end{equation}
where
\begin{equation}
  \label{eq:14}
  W_{\mu}=\sum_{k=0}^{n+\epsilon-1}\Psi_k(-x_{\mu}^2)^{n-1-k},\quad
  V_{\mu}=\sum_{k=0}^{n+\epsilon-1}\kappa_k(-x_{\mu}^2)^{n-1-k}\, , 
\end{equation}
and $\kappa_k$ and $\Psi_k$ are separation constants. For more
details, see \cite{Frolov2009a}.

We shall focus in the $D=4$ case for the rest of the paper. In this
case, we choose coordinates $(x_{1},x_{2},\psi_0, \psi_1)$, where $x^\mu$,
$\mu=1,2$, represent the PCKY eigenvalues and $\psi_i$, $i=0,1$, are
the Killing parameters of the 2 associated Killing vectors. Now if we
set $(x_1, x_2, \psi_0, \psi_1) \equiv (p, ir, t,\phi)$, the metric
\eqref{eq:metricPCKY} is written as
\begin{equation}
  \label{eq:1}
  \begin{split}
    ds^2 &= \frac{r^2+p^2}{P(p)}dp^2 + \frac{r^2+p^2}{Q(r)}dr^2 \\
    &+ \frac{P(p)}{r^2+p^2}(dt-r^2d\phi)^2 \\&-
    \frac{Q(r)}{r^2+p^2}(dt+p^2d\phi)^2\;,
  \end{split}
\end{equation}
where $P(p)$ and $Q(r)$ are $4^{\text{th}}$ order
polynomials given by \cite{Griffiths2007a}
\begin{subequations}
\begin{gather}
   \label{eq:KerrNUTAdS_polynomials_a}
   P(p) = -\frac{\Lambda}{3}p^4 -\epsilon p^2 + 2np +k,\\
   \label{eq:KerrNUTAdS_polynomials_b}
   Q(r) = -\frac{\Lambda}{3}r^4 +\epsilon r^2 - 2Mr +k,\\
   \label{eq:KerrNUTAdS_polynomials_c}
   \epsilon = 1 - (a^2+6b^2)\frac{\Lambda}{3}\;,\quad
   k = (a^2-b^2)(1-b^2\Lambda),\quad
   n = b \left[ 1+(a^2-4b^2)\frac{\Lambda}{3} \right].
 \end{gather}
\end{subequations}
The parameters are the black hole mass $M$, angular momentum to mass
ratio $a$, cosmological constant $\Lambda$, and the NUT parameter
$b$. To make contact with the physically meaningful Kerr-NUT-(A)dS
metric, we set $p=b+a\cos\theta$, $\chi^2 = 1+\Lambda a^2/3$, and make
the substitution $\phi \rightarrow \phi/a\chi^2$ and $t\rightarrow (t -
\frac{(a+b)^2}{a}\phi)/\chi^{2}$, in this order. If we set $b=0$ after
this, we have the usual Kerr-(A)dS metric in Boyer-Lindquist
coordinates \cite{Giammatteo:2005vu,Griffiths2007a}.

\subsection{Kerr-NUT-(A)dS case}
\label{sec:kerr-nut}

Let $\psi(t,\phi,r,\theta) = e^{-i\omega t}e^{im\phi}R(r)S(\theta)$ be
a solution of the Klein-Gordon equation for $D=4$ Kerr-NUT-(A)dS in
Boyer-Lindquist coordinates. The radial equation resulting from this
solution is
\begin{gather}
  \label{eq:kgAdS-4d}
  \partial_r(Q(r)\partial_r R(r)) + \left(V_{r}(r) +
    \frac{W_{r}^2}{Q(r)}\right) R(r) = 0\;,
\end{gather}
where
\begin{subequations}
\begin{gather}
  \label{eq:kerr-ads-polynomial}
  Q(r) = -\frac{\Lambda}{3}r^4 + \epsilon r^2 - 2Mr + k,\\[7pt]
  \epsilon = 1 - (a^2+6b^2)\frac{\Lambda}{3}\;,\quad
   k = (a^2-b^2)(1-b^2\Lambda),
\end{gather}  
\end{subequations}
and
\begin{subequations}
  \begin{gather}
  \label{eq:kgAdS-4d-coefficients}
  V_{r} = \kappa_0 r^2 + \kappa_1,\quad W_{r} = \Psi_0 r^2 + \Psi_1,\\[5pt]
  \kappa_0 = -4\Lambda\xi,\quad \kappa_1 = -C_l,\\[5pt]
  \Psi_0 = \omega\left(1+\frac{\Lambda a^2}{3}\right),\quad
  \Psi_1=a\left(\omega
    \frac{(a+b)^2}{a}-m\right)\left(1+\frac{\Lambda
      a^2}{3}\right)\;.
\end{gather}
\end{subequations}
The parameter $\xi$ is the coupling constant between the scalar field
and the Ricci scalar. Typical values of the parameter $\xi$ are
minimal coupling $\xi=0$ and conformal coupling $\xi=1/6$. The
separation constant between the angular and radial equations is
$C_l$. The angular equation has essentially the same form as the
radial one, associated to the problem of finding the eigenvalues of a
second order differential operator with four regular singular points,
in which case correspond to unphysical values for the latitude
coordinates. The value of its eigenvalue $C_l$ can be approximated
numerically from rational functions (Padé approximants), which were
studied in this context by \cite{Giammatteo:2005vu,Castro2013b}. We
refer to these authors for particular applications. For large values
of the energy, it can be approximated by prolate spheroidal wave
functions, whose behavior is tabulated.

In the following, we assume that all roots of $Q(r)$ are distinct and
there are two real roots at least. When $\Lambda \rightarrow 0$, two
of those roots match the Kerr horizons $(r_+,r_-)$ and the other two
diverge, leaving us with an irregular singular point of index 1 at
infinity. The characteristic coefficients -- solutions for the
indicial equations -- of the finite singularities $r_{i}$ are
\begin{equation}
  \label{eq:frobenius_coeffs}
  \rho^{\pm}_{i} = \pm i \left( \frac{\Psi_0 r_i^{2} +
      \Psi_1}{Q^{\prime}(r_i)}\right),\quad i=1,...,4
\end{equation}
and for $r=\infty$ we have 
\begin{equation}
  \label{eq:4}
  \rho_{\infty}^{\pm} = \frac{3}{2}\pm\frac{1}{2} \sqrt{9-48\xi}.
\end{equation}
These coefficients give the local asymptotic behaviour of waves
approaching any of the singular points, for example, one of the black
hole horizons. 

In this form, equation \eqref{eq:kgAdS-4d} has 5 regular singular
points, including the point at infinity. It is possible to show that
the point at infinity is actually an apparent singularity when $\xi =
1/6$ and can be further removed by a gauge transformation. In that
case, \eqref{eq:kgAdS-4d} can be cast into a Heun type equation with 4
regular singular points given by the roots of $Q(r) =
-\frac{\Lambda}{3}\prod_{i=1}^{4}(r-r_i)$. This is done in the next
section. A similar result has been reported by \cite{Suzuki:1998vy}
for massless perturbations of spin $s=0,\half,1,\frac{3}{2},2$
for Kerr-(A)dS (the so-called Teukolsky master equation) and for
$s=0,\half$ for Kerr-Newman-(A)dS. One can show that Teukolsky master
equation reduces to conformally coupled Klein-Gordon equation for
scalar perturbations, being those perturbations of the Weyl
tensor. Our computation below show this for
the spin zero case, because of the explicit non-minimal coupling, and
can be straightforwardly extended for higher spin cases. The reduction
to Heun has also been shown true for $s=\half,1,2$ perturbations of
all type-D metrics with cosmological constant \cite{Batic2007}.

\subsection{Heun equation from Conformally Coupled
Kerr-NUT-(A)dS}\label{sec:high-dimens-diff}

For $\xi=1/6$, it is possible to transform \eqref{eq:kgAdS-4d} with 5
regular singular points into a Heun equation with only 4 regular
points.  This is because $r=\infty$ in \eqref{eq:kgAdS-4d}
becomes a removable singularity. In this section, we apply the
transformations used in \cite{Batic2007} for a scalar field, adapting
the notation for our purposes\footnote{Notice that \cite{Batic2007}
  does not refer explicitly to the scalar case in their paper. However,
  our eq.~\eqref{eq:kgAdS-4d} with $\xi =1/6$ can be obtained just by
  setting $s=0$ in eq.~(11) of \cite{Batic2007}. With respect to the
  parameters of \cite{Batic2007}, we must set $a=0$ and, in a non-trivial
  change, their term $2g_4 w^2$ must be equated to $-4\Lambda\xi r^2$ to
  obtain the non-minimally coupled case.}, and we also calculate the
difference between characteristic exponents, $\theta_{i}$, for each
canonical form we obtain. As it turns out, these exponents are more
useful for us because they are invariant under generic homographic and
homotopic transformations \cite{slavyanov2000}, which preserve the
monodromy properties of a ODE, as will be seen in section 4.

By making the homographic transformation
\begin{equation}
  \label{eq:mobiustransf}
  z = \frac{r-r_1}{r-r_4}\frac{r_2-r_4}{r_2-r_1}~,
\end{equation}
we map the singular points as
\begin{equation}
  \label{eq:mobius_singularpts}
  (r_1,r_2,r_3,r_4,\infty)\quad\mapsto\quad (0,1,t_0,\infty,z_\infty) 
\end{equation}
with
\begin{equation}
  \label{eq:mobius_constants}
  z_\infty = \frac{r_2-r_4}{r_2-r_1}~,\quad t_0 =
  \frac{r_3-r_1}{r_3-r_4}z_\infty.
\end{equation}
Typically we set the relevant points for the scattering problem to $z=0$
and $z=1$, but we can consistently choose any two points to study. We
note at this point that, for the de Sitter case, $t_0$ is a real
number, which can be taken to be between $1$ and $\infty$, whereas for the
anti-de Sitter case, it is a pure phase $|t_0|=1$. Now, we define
\begin{align}
  \sigma_{\pm}(r) &\equiv \Psi_0\,r \pm \Psi_1~,\\[5pt]
  f(r) &\equiv 4\xi\Lambda r^2 + C_{l}~,
\end{align}
and
\begin{equation}
  \label{eq:aux_c_i}
  d_i^{-1} \equiv -\frac{\Lambda}{3}\prod^{3}_{\substack{j=1 \\ j\neq
i}}(r_i - r_j) = \frac{Q^{\prime}(r_i)}{r_i-r_4}.
\end{equation}
Then, eq. \eqref{eq:kgAdS-4d} transforms to
\begin{subequations}
\label{eq:kgAdS-4d_canonicalform}
  \begin{gather}
  \frac{d^2R}{dz^2} + p(z)\frac{dR}{dz} + q(z)R =0,\\[10pt]
  p(z) = \frac{1}{z}+ \frac{1}{z-1}+ \frac{1}{z-t_0} -
  \frac{2}{z-z_\infty},\\[10pt]
 q(z) = \frac{F_1}{z^2}+\frac{F_2}{(z-1)^2}+
  \frac{F_3}{(z-t_0)^2}+ \frac{12\xi}{(z-z_\infty)^2}
  + \frac{E_1}{z}+
  \frac{E_2}{z-1}+\frac{E_3}{z-t_0} + \frac{E_\infty}{z-z_\infty}, 
\end{gather}
\end{subequations}
where
\begin{subequations}
  \begin{gather}
  F_i = \left( \frac{d_i\sigma_{+}(r_i^{2})}{r_i-r_4}\right)^2 =
  \left( \frac{\Psi_0 r_i^{2} + \Psi_1 }{Q^{\prime}(r_i)}\right)^2,\\[10pt]
E_\infty = \frac{12\xi}{z_\infty(r_4 - r_1)} \left(
    \sum_{i=1}^{3}r_i - r_4 \right), \\[10pt] 
E_{i}= -\frac{d_{i}}{z_{i}-z_{\infty}}
  \left\{
    f(r_{i}) + \frac{2\Lambda}{3}\frac{d_{i}^{2}}{r_{i}-r_{4}}\sigma_{+}(r_{i}^{2})
    \left[
      \sigma_{-}(r_{i}^{2})\sum_{j\neq i}^{3}r_{j} - 2r_{i}\sigma_{-}
      \left(
         \prod ^{3}_{j\neq i}r_j
      \right)
    \right]
  \right\}~.  
\end{gather}
\end{subequations}
The $\theta_{i}$ for the finite singularities $z_i=\{0,1,t_0\}$ can be
obtained by plugging $R(z) \sim (z-z_{i})^{\theta_{i}/2}$ into
\eqref{eq:kgAdS-4d_canonicalform}
\begin{equation}
  \label{eq:ch_exp_finite_canonical}
  \theta_{i} = 2\sqrt{-F_i} = 2i \left( \frac{\Psi_0 r_i^{2} +
      \Psi_1}{Q^{\prime}(r_i)}\right)~.
\end{equation}
For $z=z_\infty$, we have that 
\begin{equation}
  \label{eq:2}
  \theta_{z_\infty} = \sqrt{9-48\xi}.
\end{equation}
These results are trivial once one knows that they are preserved
under homographic and homotopic transformations. The only remaining
singularity is $z=\infty$, in which
\begin{align}
  \label{eq:ch_exp_infinity_canonical}
  \theta_\infty &= 2i\,\sqrt{12\xi + E_2 + t_0E_3
    + z_\infty E_\infty -\sum_{i=1}^{3}\frac{\theta^2_i}{4}} \\[10pt]
&= 2i \left( \frac{\Psi_0 r_4^{2} + \Psi_1}{Q^{\prime}(r_4)}\right),
\end{align}
where the last equality follows from the invariance of $\theta$ under
homographic transformations. An important identity is that
\begin{equation}
  \label{eq:5}
 \sum_{i=1}^{4}\theta_{i} = \sum_{i=1}^{3}\theta_{i}+\theta_{\infty} = 0,
\end{equation}
by means of the residue theorem.

When the difference of any two characteristic exponents is an integer,
we have a \textit{resonant} singularity. This happens for $\xi = \{
0,5/48,1/6,3/16\}$. Thus, we have a logarithmic behaviour near
$z_\infty$, except for $\xi=1/6$ because, in this case, it is also a
removable singularity, as will be seen briefly. The property of being
removable only happens if $\theta_{z_\infty}$ is an integer different from
zero. If $\theta_{z_\infty}=0$, we always have a logarithmic singularity.
For more on this subject, see
\cite{Iwasaki:1991,slavyanov2000,Shanin:2002a}. 

To finish this section, we now show that \eqref{eq:kgAdS-4d_canonicalform}
can be transformed into a Heun equation when $\xi=1/6$. First, we make the
homotopic transformation 
\begin{equation}
  \label{eq:s_homotopic}
  R(z) =
z^{-\theta_0/2}(z-1)^{-\theta_1/2}(z-t_0)^{-\theta_t/2}(z-z_\infty)^{\beta
} \varphi(z).
\end{equation}
The transformed ODE is now given by
\begin{equation}
  \label{eq:heun_ODE}
  \frac{d^2\varphi}{dz^2} + \hat{p}(z)\frac{d\varphi}{dz} +
\hat{q}(z)\varphi =0,
\end{equation}
where
\begin{gather}
  \label{eq:heun_aux_fncts1}
  \hat{p}(z) = \frac{1-\theta_0}{z}+ \frac{1-\theta_1}{z-1}+
\frac{1-\theta_t}{z-t_0} + \frac{2\beta - 2}{z-z_\infty},\\[5pt]
  \label{eq:heun_aux_fncts2}
\hat{q}(z) = \frac{\hat E_1}{z}+
  \frac{\hat E_2}{z-1}+\frac{\hat E_3}{z-t} + \frac{\hat
E_\infty}{z-z_\infty}+ \frac{\hat F_\infty}{(z-z_\infty)^2},
\end{gather}
with
\begin{subequations}
  \begin{gather}
\label{eq:heun_E_hat_terms}
 \hat E_{i} = -\frac{d_{i}}{z_{i}-z_{\infty}}
   f(r_{i}) + \sum_{j\neq i}^{3}\frac{\theta_{i}+\theta_{j}}{2(z_{j}-z_{i})}   
   + \frac{\theta_{i}(1-\beta)+\beta}{z_{i}-z_{\infty}}~,\\[10pt]
  \label{eq:E_hats_heun}
   \hat E_\infty = \frac{12\xi}{z_\infty(r_4-r_1)}
  \left(
    \sum_{i=1}^{3}r_i - r_{4} \right) - \sum_{i=1}^{3}
    \frac{\theta_{i}(1-\beta)+\beta}{z_{i}-z_{\infty}} 
  ~,\\[10pt]
  \label{eq:F_hat_heun}
    \hat F_\infty = \beta^2 -3\beta +12\xi~.
\end{gather}
\end{subequations}

Note that $\hat F_{\infty} =0$ is the indicial polynomial associated
with the expansion at $z=z_{\infty}$.  Thus it is natural to choose
$\beta$ to be one of the characteristic exponents setting $\hat
F_{\infty} =0$. However, to completely remove $z=z_{\infty}$ from
\eqref{eq:heun_ODE}, we need that $\beta=1$ in
\eqref{eq:heun_aux_fncts1}. This further constraints $\xi =1/6$
because of \eqref{eq:F_hat_heun}. Now, we still need to check that
$\hat E_{\infty}$ can be set to zero.

The coefficients $\hat E$ above are further simplified
by noticing that
\begin{equation}
  \label{eq:3}
    \sum_{i=1}^{4}\theta_{i}r_{i} = \frac{6i\Psi_{0}}{\Lambda},\qquad 
    \sum_{j\neq i}^{3}\frac{\theta_{i}+\theta_{j}}{2(z_{j}-z_{i})}=
     - 2i\,\left(\frac{c_{i}}{z_{i}-z_{\infty}}\right)
   \left(
     \frac{\Psi_{0}r_{4}r_{i}+\Psi_{1}}{r_{i}-r_{4}}
   \right), 
\end{equation}
where again we used the residue theorem. This implies that,
for $\beta = 1$ and $\xi = 1/6$,
\begin{gather}
  \label{eq:7}
  \hat E_{i} = -\frac{d_{i}}{z_{i}-z_{\infty}}
  \left[
    f(r_{i}) + 2i
    \left(
      \frac{\Psi_{0}r_{4}r_{i}+\Psi_{1}}{r_{i}-r_{4}}
   \right)
  \right] +
  \frac{1}{z_{i}-z_{\infty}}~,\\[10pt]
 \hat E_{\infty} = \frac{1}{(r_{4}-r_{1})z_{\infty}}
 \; \sum_{i=1}^{4}r_{i}~.
 \end{gather}
The polynomial \eqref{eq:KerrNUTAdS_polynomials_b} has no third-order
term, so this means that the sum of all of its roots is zero. Therefore,
$\hat E_\infty = 0$ generically if $\beta=1$. This completes our proof
that \eqref{eq:heun_ODE} is a Fuchsian equation with 4 regular
singular points, also called Heun equation.

Summing up, the radial equation of scalar perturbations of 
Kerr-NUT-(A)dS black hole can be cast as a Heun equation in canonical form 
\begin{equation}
  \label{eq:heun_canonical}
  y^{\prime\prime} +
  \left(\frac{1-\theta_0}{z} + \frac{1-\theta_1}{z-1} +
\frac{1-\theta_t}{z-t_0} \right) y^{\prime} +
\left(\frac{\kappa_1\kappa_2}{z(z-1)}-\frac{t_0(t_0-1)K_0}{z(z-1)(z-t_0)}
\right) y = 0,
\end{equation}
with coefficients 
\begin{gather}
  \label{eq:8}
  \theta_{i} = 2i \left( \frac{\Psi_0 r_i^{2} +
      \Psi_1}{Q^{\prime}(r_i)}\right), \quad\quad K_0=-E_3,
  \quad\quad t_0=\frac{r_3-r_1}{r_3-r_4}\frac{r_2-r_4}{r_2-r_1}.
\end{gather}
The values of $\theta_i$ obey Fuchs relation, fixing $\kappa_{1,2}$
via $\theta_0+\theta_1+\theta_t+\kappa_1+\kappa_2=2$ and $\kappa_{2} -
\kappa_{1} = \theta_{\infty}$. Also, in terms of \eqref{eq:7} we have
that $\kappa_{1}\kappa_{2} = E_{2}+ t_{0}E_{3}$. These follow from the
regularity condition at infinity, $\sum_{i=1}^{3}\hat E_{i} = 0$. The
set of 7 parameters
$(\theta_0,\theta_1,\theta_t,\kappa_1,\kappa_2;t_0,K_0)$ define the
Heun equation and its fundamental solutions. By Fuchs relation, we see
that the minimal defining set has 6 parameters.

In the Kerr-NUT-(A)dS case, we note the importance of $K_0$
indexing the solutions because the only dependence on $\lambda_l$ comes
from it. As mentioned before, the local Frobenius behaviour of the
solutions do not depend on $l$, but this dependence will come about in
the parametrization of the monodromy group done below. 

The appearance of the extra singularity $t_0$ in
\eqref{eq:heun_canonical} makes things more complicated than the
hypergeometric case. First, the coefficients of the series solution obey a
three-term recurrence relation, which is not easily tractable to find
explicit solutions \cite{choun2013}.
Second, there is no known integral representation of Heun functions in
terms of elementary functions, which hinders a direct treatment of the
monodromies. Therefore, we need to look for an alternative approach to
solve the connection problem of Heun equation. In the next sections,
we will use the isomonodromic deformation theory
\cite{Jimbo1981b,Jimbo:1981-2,Jimbo:1981-3} to shed light on this
problem. 

\section{Scattering, Isomonodromy and Painlevé VI}
\label{sec:scatt-isom}

Scattering problems typically involve the calculation of a change of
basis matrix between ingoing and outgoing Frobenius solutions of two
singular points of an ordinary differential equation. This is the
connection problem of a Fuchsian differential equation, as pointed out
by Riemann and Poincaré. Fuchsian equations with 3 regular singular
points have their connection problem solved, since the solutions are
known to be expressed in terms of Gauss' hypergeometric function. For
4 regular points or more, the problem is still open. One alternative
approach to the direct computation is the study of the symmetries --
the integrable structure -- of such systems. These go by the name of
isomonodromic deformations
\cite{Jimbo:1981-2,Jimbo:1981-3,Jimbo1981b,Iwasaki:1991}. For
4 regular singular points, these are known to reduce to the study of
Painlevé transcendents, and many results about the latter came about
from the study of this integrable structure
\cite{Jimbo:1982,Okamoto:1986,Guzzetti2012}. In the following sections
we outline the application of these techniques to solve the scattering
of scalar fields around black holes.  

Linear ordinary differential equations like \eqref{eq:heun_canonical}
are of Fuchsian type because their singular points $\{0,1,t,\infty\}$
are regular: the solution behaves as $y_\pm(z) \approx
(z-z_i)^{\rho_{i}^\pm}$ near a singular point $z_i$ and then its
monodromy around each singular point is well known. By considering a
solution of either type, we have that $y_\pm(e^{2\pi
  i}(z-z_i))=e^{2\pi i\rho^{\pm}_i}y_\pm(z)$. In the following, we
suppose that $\rho^{\pm}$ are different, finite, and non-zero complex
numbers whose difference is not an integer. The most natural setup to
study monodromies are Fuchsian systems because, as mentioned in the
introduction, the number of parameters defining them match the number
of parameters of monodromy representations. Any Fuchsian equation can
be written as a linear Fuchsian system with an appropriate gauge
connection $A(z)$,
\be
\Dp\mathcal{Y}(z)=A(z)\mathcal{Y}(z),
\label{eq:schlesinger}
\ee
where $\mathcal{Y}(z)$ is a column vector of two functions $y_1(z)$
and $y_2(z)$ \cite{Iwasaki:1991}. Now, let 
\begin{equation}
  \label{eq:9}
  A(z) =
  \begin{pmatrix}
    A_{11}(z) & A_{12}(z) \\
    A_{21}(z) & A_{22}(z) 
  \end{pmatrix}.
\end{equation}
It can be verified that $y_1(z)$ satisfies the equation
\be
 y'' -\left(\frac{A_{12}'}{A_{12}}+\Tr A(z)\right)
 y'+\left(\det A(z)-A_{11}'+A_{11}\frac{A_{12}'}{A_{12}} 
 \right)y=0,
\label{eq:garnier_edo}
\ee
with a similar equation for $y_2(z)$. Now, if we are given a
fundamental matrix of solutions $\Phi(z)
=(\mathcal{Y}_{1}(z),\mathcal{Y}_{2}(z))^{T}$, we can write the
connection in 
terms of it 
\be
A(z)=[\Dp\Phi(z)]\Phi^{-1}(z),
\label{eq:puregauge}
\ee 
which tells us that $A = A(z)dz$ can be seen as a ``pure gauge''
$GL(2,\mathbb{C})$ gauge field, satisfying $F=dA+A\wedge A=0$. Since
we are working in the $n$-punctured Riemann sphere, 
we are free to consider gauge transformations $\Phi(z)\rightarrow
U(z)\Phi(z)$, or analogously, $A(z)\rightarrow U(z)A(z)U^{-1}(z)+\Dp
U(z)U^{-1}(z)$, where $U(z)$ has meromorphic functions for
entries. These meromorphic functions can introduce apparent
singularities, in which the indicial equation of
\eqref{eq:garnier_edo} has integer values, and 
there is no logarithmic branching point. In this case the monodromy
matrix around the apparent singularity is trivial: any composition of
loops enclosing apparent singularities will have no effect on the
monodromy associated with the loop. 

As it turns out, an apparent singularity is exactly what one has in
\eqref{eq:garnier_edo} when $A_{12}$ vanishes. Let $t
=(t_{1},t_{2},\dots,t_{n+3})$ represent a set of $n+3$ singular points
on the Riemann sphere, including $t_{n+1}=0,t_{n+2}=1$ and $t_{n+3}=\infty$,
and let $\lambda = (\lambda_{1},\lambda_{2},\dots,\lambda_{n})$
represent the zeros of $A_{12}$. A Fuchsian system of
\textit{Schlesinger type} is written in a gauge where $A(z)$ has a
partial fraction expansion 
\be 
A(z,t)dz=\sum_i\frac{A^i(t)}{z-t_i}\,dz,
\label{eq:fuchsconnection}
\ee 
where $A^{i}$ are matricial coefficients depending only on $t$. We
can now ask if there is a way to change the positions of the regular
singular points $t$ keeping the monodromies of \eqref{eq:schlesinger}
invariant. For that matter, we introduce the auxiliary system
\begin{equation}
  \label{eq:10}
  \Dt\mathcal{Y}(z,t) = B(z,t)\mathcal{Y}(z,t).
\end{equation}
As it turns out, in the Schlesinger gauge, 
\begin{equation}
  \label{eq:11}
  B(z,t)dt = - \sum_i\frac{A^i(t)}{z-t_i}\,dt_{i}\,,
\end{equation}
and the integrability condition for the Pfaffian system formed by
\eqref{eq:schlesinger} and \eqref{eq:10} is given by the so called 
Schlesinger's equations
\cite{Schlesinger:1912,Iwasaki:1991} \be \frac{\partial A^i}{\partial
  t_j}=\frac{[A^i,A^j]}{t_i-t_j}, \quad j\neq i \quad\quad \text{and}
\quad\quad \frac{\partial A^i}{\partial t_i}=-\sum_{j\neq
  i}\frac{[A^i,A^j]}{t_i-t_j}.
\label{eq:schlesinger_general}
\ee
The isomonodromy flow generated by the Schlesinger system above is
Hamiltonian and has been studied
extensively in a series of papers by Jimbo, Miwa and collaborators
\cite{Jimbo1981b,Jimbo:1981-2,Jimbo:1981-3}. For the case of interest, 
the Heun equation, the phase space is two-dimensional, as we see
below. We will use the 
asymptotics of the isomonodromy flow in order to solve for the monodromy
problem of \eqref{eq:heun_canonical}.

To better clarify the last paragraph and make contact with the
monodromy problem of \eqref{eq:heun_canonical}, we need to understand
how the isomonodromic flow act on \eqref{eq:garnier_edo} in the case
$n=1$. Let us 
choose a gauge where $\Tr A^i=\theta_i$ and where the off-diagonal
terms of $A$
decay as $z^{-2}$ as $z\goesto\infty$. Then $A_{12}(z,t)$ has a single zero
at $z=\lambda$, and is of the form:
\be
A_{12}(z,t)=\frac{k(z-\lambda)}{z(z-1)(z-t)}.
\label{eq:a12}
\ee 
We fix the asymptotic behavior:
\be
A_\infty=-(A^0+A^1+A^t)=
\begin{pmatrix} 
  \kappa_1 & 0 \\
  0 & \kappa_2-1
\end{pmatrix},
\ee 
with $\kappa_1+\kappa_2=1-\theta_0-\theta_1-\theta_t$ and
$\kappa_2-\kappa_1=\theta_\infty$ related to the parameters of the
singular points. This choice introduces an extra
singularity in \eqref{eq:garnier_edo}.

Plugging \eqref{eq:9} with \eqref{eq:a12} into \eqref{eq:garnier_edo}, we find
a Fuchsian differential equation of \textit{Garnier type}
\begin{subequations}
\label{eq:garnier}  
\begin{gather}
y''+p(z,t)y'+q(z,t)y=0, \\[10pt]
p(z,t)=\frac{1-\theta_0}{z}+\frac{1-\theta_1}{z-1}+\frac{1-\theta_t}
{
z-t}-\frac{1}{z-\lambda}, \\[10pt]
q(z,t)=\frac{\kappa_1\kappa_2}{z(z-1)}-\frac{t(t-1)K}{z(z-1)(z-t)}+
\frac{\lambda(\lambda-1)\mu}{z(z-1)(z-\lambda)}.
\end{gather}
\end{subequations}
The parameters $K$ and $\mu$ will play a significant role in the
following and are related to $A(z)$ by
\begin{gather}
\mu = \frac{A_{11}^0}{\lambda}
+\frac{A_{11}^1}{\lambda-1}+\frac{A_{11}^t}{\lambda-t}, \\[10pt]
\begin{split}
K =\frac{A^t_{11}}{\lambda-t}+ \frac{A_{11}^0+A_{11}^t-\theta_0\theta_t}{t}
&+  \frac{A_{11}^1+A_{11}^t-\theta_1\theta_t}{t-1} \\[5pt]
&+\frac{1}{t}\Tr A^0A^t+\frac{1}{t-1}\Tr A^1A^t.
\end{split}
\end{gather}
In order to ensure that the singularity at $z=\lambda$ is apparent,
$K$ is constrained to be a specific rational function of $\mu$, $\lambda$ and
$t$,
\be
K(\lambda,\mu,t)=\frac{\lambda(\lambda-1)(\lambda-t)}{t(t-1)}
\left[\mu^2-\left(\frac{\theta_0}{\lambda}+\frac{\theta_1}{\lambda-1}+
\frac{\theta_t-1}{\lambda-t}\right)\mu+\frac{\kappa_1\kappa_2
}{\lambda(\lambda-1)}\right].
\label{eq:painleve6h}
\ee
The interesting thing about writing \eqref{eq:garnier} in this form is
that $K$ is a hamiltonian generating its isomonodromic flow of
 in terms of $(\lambda(t),\mu(t))$. The isomodromic
flow shuffles around the position of the apparent singularity $\lambda$
and the ``conjugate momentum'' $\mu$ in such a way to keep the monodromies
fixed. Specifically, a change in the position of the true singularity
$t$ entails a change in the parameters given by the \textit{Garnier system}
\begin{equation}
  \Dfrac{\lambda}{t} =\{K,\lambda\},\quad\quad
  \Dfrac{\mu}{t}= \{K,\mu\},
\label{eq:hamiltonianstructure}
\end{equation}
where the Poisson bracket is defined by
\be
\{f,g\}=\frac{\partial f}{\partial \mu}\frac{\partial g}{\partial \lambda}
-\frac{\partial f}{\partial \mu}\frac{\partial g}{\partial \lambda}.
\ee
The Schlesinger equations \eqref{eq:schlesinger_general} are given in
this case by (see, for instance, \cite{Hitchin:1997zz}):
\be
\frac{dA^0}{dt}=\frac{[A^t,A^0]}{t},\quad\quad
\frac{dA^1}{dt}=\frac{[A^t,A^1]}{t-1},\quad\quad
\frac{dA^t}{dt}=\frac{[A^0,A^t]}{t}+\frac{[A^1,A^t]}{t-1}.
\label{eq:schlesinger_heun}
\ee 
The Hamiltonian associated with the Schlesinger system is
$t(t-1)H=(t-1)\Tr A^0A^t+t\Tr A^1A^t$, which is the term in the second
line of the expression for $K$ above. The two Hamiltonians $K$ and $H$
are thus related by a canonical transformation
\cite{Okamoto:1986,Iwasaki:1991}. Since the entries $A^0_{11}$,
$A^1_{11}$ and $A^t_{11}$ can be explicitly computed in terms of
$\mu,\lambda,t$ the Garnier and the Schlesinger systems are actually
equivalent. Explicit expressions can be found in
\cite{Iwasaki:1991,Jimbo:1981-2}. 

Our conclusion is that $\mu$ and $\lambda$ are canonically conjugated
coordinates in the phase space of isomonodromic deformations. If we
write the equation of motion in terms of $\lambda$ alone,
\begin{multline}
\ddot{\lambda}= 
\frac{1}{2}\left(\frac{1}{\lambda}+\right.
\left.\frac{1}{\lambda-1}+\frac{1}{\lambda-t}\right)
\dot{\lambda}^2-\left(\frac{1}{t}+\frac{1}{t-1}+\frac{1}{\lambda-t}
\right)\dot{\lambda} 
\\
+\frac{\lambda(\lambda-1)(\lambda-t)}{2t^2(1-t)^2}
\left(\theta_\infty^2-\theta_0^2\frac{t}{\lambda^2}+
\theta_1^2\frac{t-1}{(\lambda-1)^2}+
\left(1-\theta_t^2\right)\frac{t(t-1)}{(\lambda-t)^2}\right),
\end{multline}
which corresponds to the sixth Painlevé equation $P_{VI}$.  This is
the more general second order differential equation of the form
$\ddot{z}=R(z,\dot{z},t)$, with $R$ a rational function, which has the
Painlevé property: the singularities of $\lambda(t)$, apart from
$t=0,1,\infty$, are simple poles and depend on the choice of initial
conditions. Given a particular set of initial conditions, the equation
can then be used to define a new transcendental function, the Painlevé
transcendent ${\cal
  P}_{VI}(\theta_\infty,\theta_0,\theta_1,\theta_t;t)$, in the same
way the linear second order ordinary equation with $3$ regular
singular points can be used to define the hypergeometric function
\cite{Iwasaki:1991,Guzzetti2012}. 

Now we see how the theory of isomonodromic deformations can help us to
solve our initial scattering problem: Painlevé VI asymptotics are
given in terms of the monodromy data of \eqref{eq:garnier}. In section
5, we show how to relate \eqref{eq:garnier} with
\eqref{eq:heun_canonical} and how Painlevé asymptotics solve the
monodromy problem of Heun equation. But first, in the next section, we
make a mathematical digression about how to parameterize the monodromy
group of Fuchsian systems.

\section{Flat Connections and Monodromies}

Physically, the formulation in terms of the flat connection
\eqref{eq:puregauge} with the decomposition \eqref{eq:fuchsconnection}
means that the scattering problem is equivalent to
finding the potential of a holomorphic $GL(2,\mathbb{C})$ Yang-Mills field
with a number of monopoles with non-abelian charges $A^i$
\cite{Atiyah:1982}. It is then reasonable to expect that, given the
positions of the charges and their coefficients, the monodromy values
will be uniquely defined.

Mathematically, the space of such flat connections, ${\cal A}_{g,n}$, is
associated with the moduli space of genus $g$ Riemann surfaces with $n$
punctures ${\cal M}_{g,n}$, and our case of interest will be the Riemann
sphere where $g=0$. Because of global conformal transformations,
\be
z\rightarrow \tilde{z}=\frac{az+b}{cz+d},\quad\quad
\tilde{A}(\tilde{z})d\tilde{z}=A(z)dz,
\ee
we are able to fix $3$ of the $n$ singular points and ${\cal M}_{g,n}$
is thus covered
by the anharmonic ratios like
\be
\frac{(z-z_n)(z_{n-1}-z_{n-2})}{(z-z_{n-2})(z_{n-1}-z_{n})},
\ee
which allowed us to write Heun equation \eqref{eq:heun_canonical} in
terms of the position of a single pole $t$ alone. The other
independent parameter
appearing in \eqref{eq:heun_canonical} is $K_{0}$. In order to
describe it geometrically, let us 
consider a generic Fuchsian equation in the normal form with $n$
finite singular points, that is  
\begin{gather}
  \label{eq:heun-normal-finite}
  \psi^{\prime\prime}(w) + T(w)\psi(w) = 0,\qquad T(w) = \sum_{i=1}^{n}
  \left(
  \frac{\delta_i}{(w-w_i)^2} + \frac{c_i}{w-w_i}
  \right),\\[10pt]
  \sum_{i=1}^{n} c_i = 0\;,\quad \sum_{i=1}^{n}( c_iw_i +\delta_i) =
  0\;,\quad \sum_{i=1}^{n}( c_iw^2_i +2\delta_iw_i) =
  0  \label{eq:heun-normal-finite2}, 
\end{gather}
where \eqref{eq:heun-normal-finite2} are the necessary and sufficient
conditions for $w=\infty$ to be a regular point. Because of
\eqref{eq:heun-normal-finite2} there are only $n-3$ independent
$c_{i}$, and we can also fix 3 of the $z_{i}$ to be $0,1$ and $\infty$
by a homographic transformation. Thus, if we fix the $\delta_{i}$, we
can parametrize Fuchsian equations by $2(n-3)$ complex numbers
$(c_{i},z_{i})$. Typically we say that $\delta_{i}$ and $z_{i}$ are
local parameters, depending only on local behaviour of solutions, and
the $c_{i}$, usually called accessory parameters, have global
properties not probed locally. The accessory parameters are usually
related to spectral parameters of differential equations
\cite{ronveaux1995heun,slavyanov2000}. We notice now that the angular
eigenvalue $\lambda_{l}$ dependence appears exactly in the accessory
parameter of Heun equation, and that is why it did not appear in the
Frobenius coefficients $\theta_{i}$. 

We can relate \eqref{eq:heun-normal-finite} to our Heun equation
\eqref{eq:heun_canonical} by setting $n=4$ and applying a homographic
transformation such that $(w_1,w_2,w_3,w_4,\infty;w) \mapsto
(0,1,t,\infty,z_\infty;z) $. We also need to remove the apparent
singularity by letting $\psi \mapsto (z-z_\infty)^{-1}\psi$, finally
giving us the equation
\begin{equation}
  \label{eq:heun-normal-litvinov}
  \psi^{\prime\prime}(z) + \tilde T(z)\psi(z) = 0,\qquad \tilde T(z) =
  \sum_{i=1}^{3} 
  \left(
  \frac{\delta_i}{(z-z_i)^2} + \frac{\tilde c_i}{z-z_i}
  \right)\;,
\end{equation}
such  that
\begin{equation}
  \label{eq:aux-normal-litvinov}
  \tilde c_i =
  \frac{c_i(w_{4}-w_{i})-2\delta_i}{z_{i}-z_{\infty}}\;,\qquad
  \sum_{i=1}^{3} \tilde c_i = 0. 
\end{equation}
Note that 
\begin{equation}
  \label{eq:19}
  \fosum{c_iw_{4i}^2-2\delta_iw_{4i}} = \frac{\tilde c_2 + t\tilde
    c_3}{z(z-1)} + \frac{t(t-1)\tilde c_3}{z(z-1)(z-t)}\;. 
\end{equation}
Analyzing the behaviour of \eqref{eq:heun-normal-litvinov} at
infinity, we may rewrite $\tilde c_2 + t\tilde c_3 $ as $\delta_4 -
(\delta_1 + \delta_2+\delta_3)$. Finally, we now take
\eqref{eq:heun_canonical} and make $y(z) =
N(z)\psi(z)$, where 
\begin{equation}
  \label{eq:16}
  N(z) = \exp\left(-\int \hat p(z) dz\right) =
  \prod_{i=1}^{3}(z-z_{i})^{-(1-\theta_{i})}, 
\end{equation}
to obtain \eqref{eq:heun-normal-litvinov}. Within this transformation, we
can check that
\begin{align}
  \label{eq:17}
\tilde T(z) &= \hat q(z) - \frac{\hat{p}^2(z)}{4} -
\frac{\hat{p}'(z)}{2},   
\end{align}
which implies  $\delta_{i} = (1-\theta_{i}^{2})/4$ and also
\begin{equation}
  \label{eq:18}
  \tilde c_{i} = \frac{1-d_{i}f(r_{i})}{z_{i}-z_{\infty}} +
  \sum_{j\neq i}^{3} \frac{(1+\theta_{i}\theta_{j})}{2z_{ji}}.
\end{equation}

One of the most important results of the mathematical investigation in
\cite{Atiyah:1982} is that the space of flat connections ${\cal
A}_{0,n}$ has a natural symplectic form $\Omega$. It basically stems from
the fact that flat connections have a natural action, the Chern-Simons
form, living in a space with an extra dimension, apart from the
coordinates $z,\bar{z}$, with the extra dimension interpreted as a gauge
parameter:
\be
\delta S=\int_{\Sigma\times \mathbb{C}}\Tr(\delta A\wedge
F)+2\int_{\Sigma}\Tr(\delta A\wedge A),
\ee
Thus one relates the variation of the boundary
``$p\delta q$'' term, $\frac{1}{2\pi i}\Tr(\delta A\wedge \delta A)$, to
the Atiyah-Bott symplectic form $\Omega$. The construction is reminiscent
of the appearance of the so-called Wess-Zumino term in WZW models in
conformal field theory. As it turns out
\cite{Krichever:2007cx,Nekrasov:2011bc}, the $n-3$
independent accessory parameters and the $n-3$ independent poles $z_i$ are
a set of Darboux coordinates for $\Omega$, that is:
\be
\Omega=\sum_i^{n-3}dc_i\wedge dz_i.
\ee
Specializing to the four singularity case, we have that, in terms of the
canonical form of the equation \eqref{eq:heun_canonical}, the symplectic 
form can be readily written in terms of the position of the singularities
and the ``Hamiltonian'': 
\be
\Omega=dK\wedge dt,
\ee
as can be anticipated from  the Hamiltonian form of the Painlevé
equation. One has a heuristical correspondence between Heun's equation
and  Painlevé VI: the latter can be understood as the classical
hamiltonian system \eqref{eq:hamiltonianstructure}, whereas Heun's
equation is obtained from $K$ if we treat it as the ``quantum
hamiltonian''. In fact, from \eqref{eq:painleve6h} one has:
\be
H\left(z,-\frac{\partial}{\partial z},t\right)y(z)=Ky(z)
\ee
as the Heun equation, modulo an integer shift of the
$\theta_i$\footnote{This was called a Schlesinger transformation in 
\cite{Jimbo1981b,Jimbo:1981-2,Jimbo:1981-3}. Note that the integer shift
doesn't change monodromies around a single singular point.}. Thus, in a
sense, the Painlevé VI equation is the classical limit of the
Heun equation \cite{Slavyanov:1996}. This also reinforces the view that
the position of the singular point $t$ and the accessory parameter $K$
should be viewed as conjugate quantities. 

Incidentally, the other five equations from the Painlevé list can be
obtained from the sixth by a scaling limit (confluence). An object of
further study is whether this allows for calculation of monodromies in
the confluent Heun case.

The flow means that the ``phase space'' $\{z_n,c_n\}$ can be foliated
into integral curves of the Hamiltonian \eqref{eq:painleve6h}. The
monodromy data is constant over these curves, so they are effectively
functions of the space of orbits. We want, however, to get the monodromy
data from the values of $h$ and $t$. In order to do this, we will have to
review the algebraic aspects of the monodromy matrices.

The Fuchsian equation is defined in the Riemann sphere
$\mathbb{CP}^1\sim \mathbb{C}\cup \{\infty\}$ minus $n$ points
$\{z_1,z_2,\ldots, z_n\}$ in which the fundamental matrix of solutions
diverge like $(z-z_i)^{\alpha^\pm_i}$ with $\alpha^\pm_i$ the solutions
of the indicial equation. Following the usual Riemann-Hilbert problem
formulation \cite{Anosov1994}, we will define the fundamental group of
such space with a fixed point $z_0$, and construct a representation as
follows. Let $\gamma_i$ be a curve containing $z_0$ that divides the
punctured sphere into two regions, one containing only $z_i$ and the
other containing all other singular points. We associate with $\gamma_i$
a matrix $M_i$ which mixes the two solutions of the general ODE
\eqref{eq:kgAdS-4d}. Clearly
\be
M_1M_2\ldots M_n=\mathbb{I},
\label{eq:monoproblem}
\ee
since the composition of all $\gamma_i$ is a contractible curve. The
famous Riemann-Hilbert problem consists in finding an ODE with a given
set of monodromy data $M_i$. Our problem is quite the opposite: how to
determine $M_i$ from the data readily available in the ODE.

Of course the problem does not have a single solution, if one finds a
particular set of matrices $\{M_i\}$ satisfying \eqref{eq:monoproblem},
then $\{gM_ig^{-1}\}$ will also be a solution, corresponding to a
diferent choice of fundamental solutions. Also, the indicial equation
allows us to write the solution near a singular point: up to a change of
basis, the monodromy matrix near a regular singular point is:
\be
M_i\sim \Lambda_i\exp[\rho_i\mathbb{I}+\alpha_i\sigma^3],\quad\quad
\rho^\pm_i=\rho_i\pm\alpha_i,
\ee
so the conjugacy class of each $M_i\in GL(2,\mathbb{C})$ is known. In
the following we will assume without loss of generality that
$\rho_i^0=0$, which reduces the group to $SL(2,\mathbb{C})$. For the
application to ODEs these are set by the Fuchs
relation. Let us define $g_i$ as the matrix that changes basis between the
fiducial point $z_0$ and the $z_i$:
\be
M_i=g_i \exp[\alpha_i\sigma^3]g_i^{-1},\quad\quad \alpha_i=2\pi i
\theta_i.
\ee
The determination of the $g_i$ is important for computing scattering
elements. Given a ``purely ingoing'' or ``purely outgoing'' solution near
$z_i$, the scattering amplitudes from another point $z_j$ are given by
\be
{\cal M}_{i\rightarrow j}=g_ig_j^{-1}=
\left(
\begin{array}{cc}
 1/{\cal T} & {\cal R}/{\cal T} \\
 {\cal R}^*/{\cal T}^* & 1/{\cal T}^*
\end{array}
\right)
\label{eq:scattering_matrix}
\ee
where ${\cal T}$ and ${\cal R}$ are the transmission and reflection
amplitudes, respectively. 

We can now turn back to the problem of relating the monodromy data to the
accessory parameters in \eqref{eq:heun_canonical}. We will review the
construction outlined in \cite{Nekrasov:2011bc}. Let $M_i$ be the
monodromy matrix as above. The monodromy data readily available from the
differential equation are the traces:
\be
m_i=\Tr(M_i),\quad i=0,1,t.\quad\quad\text{and}\quad\quad
m_{\infty}=\Tr(M_0M_1M_t)=\Tr(M_\infty^{-1}).
\ee
In order to fully characterize the $M_i$ (up to an overall conjugation),
we need the other characters:
\be
m_{01}=\Tr(M_0M_1),\quad\quad m_{0t}=\Tr(M_0M_t),\quad\quad
m_{1t}=\Tr(M_1M_t).
\ee
The set of $m$'s are not all independent, they satisfy the Fricke-Jimbo
relation:
\be
\begin{aligned}
W(m_{0t},m_{1t},m_{01})=m_{0t} & m_{1t}m_{01} 
+m_{0t}^2+m_{1t}^2+m_{01}^2
-m_{0t}(m_1m_{\infty}+m_0m_t) \\ & -m_{1t}(m_0m_{\infty}+m_1m_t) 
-m_{01}(m_tm_{\infty}+m_0m_1) \\ & \quad\quad\quad +m_0^2+m_1^2
 +m_t^2+m_{\infty}^2+ m_0m_1m_tm_{\infty}=4.
\label{eq:Fricke_relation}
\end{aligned}
\ee
Which gives a quadratic relation that allow one to compute one of the
$m_{ij}$, say, $m_{1t}$, given the other two, $m_{01}$ and
$m_{0t}$. The configuration space for the monodromy data with fixed
$m_i$ and $m_{\infty}$ is then parametrized by 2 independent variables. We
can give the solution for the monodromy matrices parametrized by the set
of $m_i$ and $m_{ij}$: Given a Euler-angle parametrization of the $g_i$'s:
\be
g_i=\exp[\psi_i\sigma^3/2]\exp[\phi_i\sigma^1/2]\exp[\varphi_i\sigma^3/2],
\ee
one notes immediately that $\varphi_i$ can all be set to zero, while the
parametrization of the $m_{ij}$ can be verified by simple matrix
multiplication:
\be
m_{ij}
=2\cosh \alpha_i\cosh\alpha_j+2\sinh\alpha_i\sinh\alpha_j(\cosh\phi_i\cosh
\phi_j+\sinh\phi_i\sinh\phi_j\cosh(\psi_i-\psi_j)).
\ee
And $m_{\infty}$ given by the Fricke-Jimbo relation. One notes that there is
an overall symmetry $\psi_i\goesto \psi_i+\psi$, which can be used to
reduce the overall number of parameters to $5$, in the Heun case. We are
given the four $\theta_i$, so there is one unfixed free parameter in the
monodromy matrices which depends explicitly on the accessory parameter $K$
and the anharmonic ratio $t$.

With the $\theta_i$ fixed, the set of two complex numbers
$m_{01}$ and $m_{0t}$ provide a local set of coordinates to the
space of flat connections ${\cal A}_{0,4}$. These coordinates are not
canonical, in the sense we will explore now.

As stated in the preceding Section, ${\cal A}_{g,n}$ has a natural
symplectic structure, given by the Atiyah-Bott formula:
\be
\Omega = \frac{1}{2\pi i}\int_\Sigma \Tr(\delta A\wedge \delta A).
\label{eq:symplectic2}
\ee
The traces of the monodromies are formally given by the Wilson loops:
\be
m_{\gamma} = \Tr M_{\gamma} = \Tr P\exp
\left[\oint_{\gamma}A(z)dz\right]. 
\ee
Using \eqref{eq:symplectic2} one can compute the skein-relations, relating
different holonomies \cite{Turaev:1991}:
\be
\left\{m_{\gamma_i},m_{\gamma_j}\right\} = \frac{1}{2}
\sum_{x\in
\gamma_i\cap\gamma_j}\left(m_{\gamma^+_{x,i,j}}-m_{\gamma^-_{x,i,j}}
\right),
\ee
where the loops $\gamma^\pm_{x,i,j}$ are constructed by removing a small
neighborhood of the intersection point $x$ and replacing it by two arcs.
The superscript labels the two choices of completion. See Figure 1.

\begin{figure}[hbt]
\begin{center}
 \mbox{\includegraphics[width=0.95\textwidth]{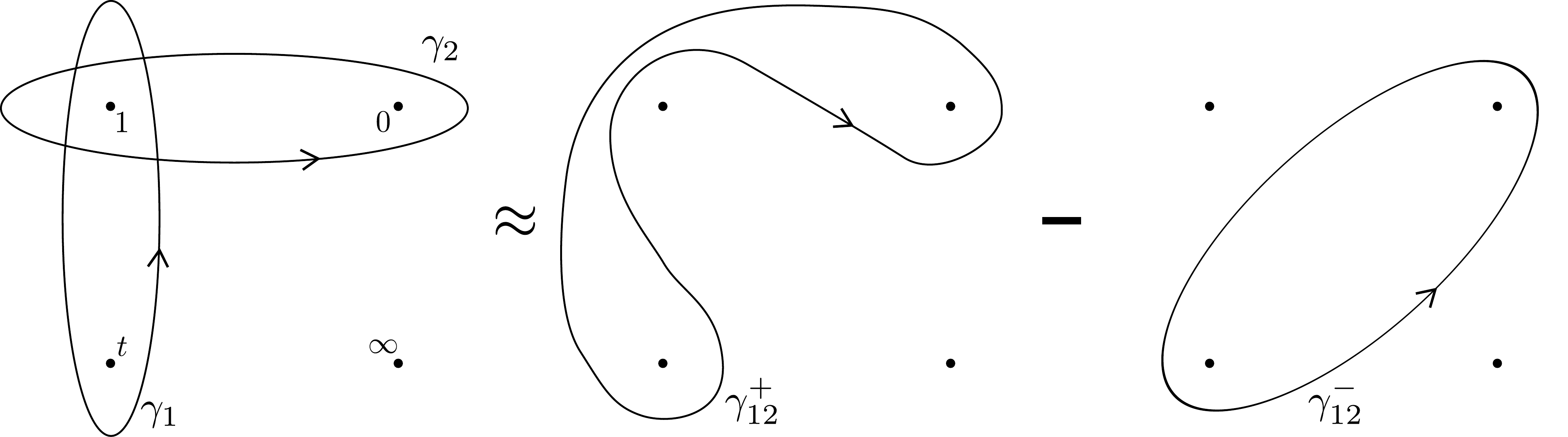}}
\end{center}
\caption{The skein-relation applied to the four-punctured sphere. In the
left hand side we represent the Poisson bracket by drawing both curves
simultaneously.}
\end{figure}

In terms of the variables $m_{01}$, $m_{0t}$ and $m_{1t}$,
the skein relation implies:
\be
\{m_{1t},m_{01}\}=\rho^+-m_{0t}
\ee
where, because of the Cayley-Hamilton theorem, we have for an unimodular
matrix $\Tr(A)\mathbb{I}=A+A^{-1}$, and then:
\be
\rho^+=\Tr(M_1^{-1}M_tM_1M_0)=-m_{0t}-m_{1t}m_{01}+m_0m_t+m_1m_{\infty}.
\ee
Note also that the Poisson bracket is related to the function $W$ defined
by \eqref{eq:Fricke_relation}:
\be
\{m_{1t},m_{01}\}=-\frac{\partial W}{\partial m_{0t}}.
\ee
Given this relation, one can now introduce canonically conjugated
coordinates $\phi$ and $\psi$ on ${\cal M}_{0,4}$, so that the
monodromies $m_{ij}$ are parametrized as follows:
\be
\begin{aligned}
 m_{1t} = & 2\cos\pi\phi, \\
 m_{01} = & \frac{2\cos\pi\psi}{m_{1t}^2-4}\sqrt{c_{1t}c_{0\infty}}
 -2\frac{(m_0m_1+m_tm_{\infty})-\cosh\pi\phi(m_0m_t+m_1m_{\infty})
 }{m_{1t}^2-4}, 
 \\ 
 m_{0t} = & \frac{\sin\pi\phi}{2\sin\pi\psi}\sqrt{c_{1t}c_{0\infty}} 
-\frac{1}{2}\left(m_{1t}m_{01}-m_0m_t-m_1m_{\infty}\right);
\end{aligned}
\ee
with
\be
c_{1t}=m_{1t}^2+m_1^2+m_t^2-m_{1t} m_0m_1-4,\quad\quad
c_{0\infty}=m_{1t}^2+m_0^2+m_{\infty}^2-m_{1t} m_0m_{\infty}-4.
\ee
Given that $\phi$ and $\psi$ are independent Darboux coordinates, we
have, up to a multiplicative constant,
\be
\Omega = d\phi \wedge d\psi.
\ee
And then the transformation from the parameters in the Heun equation
\eqref{eq:heun_canonical} $t,K$ to the monodromy parametrization
$\phi,\psi$ is canonical.  

\section{The Classical Mechanics of Monodromies}

Now we turn to the problem of finding the canonical transformation that
takes the parameters of the Heun equation to the monodromy parametrization
$\phi$ and $\psi$. By canonical, one means that there exists a function
$f$ such that
\be
K\equiv K(\phi,t)=\frac{\partial}{\partial t}f(\phi,t),
\label{eq:canonicalf}
\ee
with $f(\phi,t)$ the generating function of the transformation.
This function has been receiving some attention recently because its
relation to conformal blocks in Liouville field theory
\cite{Zamolodchikov:1987aa,Zamolodchikov:1995aa}. For
recent developments both in the application for Liouville and $c=1$
conformal blocks see
\cite{Alday:2009aq,Mironov:2011aa,Gamayun:2012ma,Litvinov:2013sxa,
Iorgov:2014vla} .
There it appears as the semiclassical approximation to the $5$-point
function of conformal primaries. It is the WKB approximation to the Ward
identity:
\be
\left[\frac{1}{b^2}\frac{\partial^2}{\partial z^2}+\sum_{n=1}^4\left(
\frac{\Delta_i}{(z-z_i)^2}+\frac{1}{z-z_i}\frac{\partial}{\partial z}
\right)\right]\langle V_{(1,2)}(z)V_{\Delta_1}(z_1)\ldots
V_{\Delta_4}(z_4)\rangle = 0,
\ee
up to contact terms. Because of conformal invariance, this expectation
value also depends only on the anharmonic ratios between the
coordinates, $\{t_i\}$, and the classical limit yields:
\be
\langle V_{(1,2)}(z)V_{\Delta_1}(z_1)\ldots
V_{\Delta_4}(z_4)\rangle_{b\rightarrow 0}=\psi(z,t)\exp\left(
{\textstyle \frac{1}{b^2}} f(\phi,t)\right).
\ee
Where $\psi(z,t)$ is a solution of the Heun equation in the normal form
and $f(\phi,t)$ is as in \eqref{eq:canonicalf}. For more details about
Liouville correlators and the Riemann-Hilbert problem see
\cite{Ginsparg:1993is}. 

Now the equations of motion for isomonodromic transformations is
obvious in terms of the variables $\phi$ and $\psi$: they are
constant. Therefore, the generating function of the canonical
transformation is the action itself, calculated at a solution of the
Painlevé equation \cite{Arnold:1989}. To wit, let us remind that, since
we are dealing with canonically conjugate coordinates, we can write the
action 1-form $\alpha$ using either pair:
\be
\alpha=\mu d\lambda - K(\mu,\lambda,t) dt = \psi d\phi - H(\phi,\psi,S)
dS.
\ee
The Hamiltonian with respect to the $\phi$ and $\psi$ coordinates is
trivial, since the flow is isomonodromic. We will take $H=-1$ and
identify $S$ with the action computed on the solutions of
the equations of motion: 
\be
dS=\mu d\lambda - K(\mu,\lambda,t) dt-\psi d\phi.
\label{eq:symplectic}
\ee
Hence
\be
\mu=\frac{\partial S}{\partial \lambda},\quad\quad \psi=-\frac{\partial
S}{\partial \phi},\quad\quad K=-\frac{\partial S}{\partial t}.
\ee
That is, $S$ is a function of the independent variables $\lambda$, $t$ and
$\phi$. Observing the last equality, we can then invert and have the
monodromy as a function of $K$ and $t$:
\be
K=-\frac{\partial S}{\partial t}(\lambda,\phi,t)\Rightarrow 
\phi=\phi(K,\lambda,t).
\ee
Now, if the system is computed at the solutions of the equation of motion,
$\lambda(t)$ satisfies the Painlevé VI and $\mu(t)$ is given by
\be
\dot{\lambda}=\frac{\partial K}{\partial
  \mu}=\frac{\lambda(\lambda-1)(\lambda-t)}{t(t-1)} 
\left[2\mu-\left(\frac{\theta_0}{\lambda}+\frac{\theta_1}{\lambda-1}
+\frac{\theta_t-1}{\lambda-t}\right)\right].
\ee
Given that $\phi$ is constant over the solutions, this leads to
\be
S(\phi,\lambda,t)=\int_{(\lambda_i,t_i)}^{(\lambda,t)}
\mu(\lambda,t') d\lambda - K(\mu(\lambda,t'),\lambda,t')dt'.
\label{eq:classical_painleve}
\ee
The path of integration $(\lambda(s),s)$ is a solution of the Painlevé VI
equation with initial condition given by $\lambda_i,t_i$ and monodromy
parameter given by $\phi$. The dependence on $\phi$ has been considered in
a number of papers \cite{Jimbo:1982,Boalch:2005}, and is explicit at the
Painlevé singular points $t=0,1,\infty$. Let us take $t_i\goesto 1$ as the
asymptotic point for definiteness. One should note that we can always take
this to be the case by a permutation of the singular points of the Heun
equation -- whose action in the Painlevé equation is known as the
bi-rational transformation \cite{Okamoto:1986}. Near the singular point,
we assume further that $0<\Re\phi<1$, so one can find that (see Appendix):
\be
\lambda(t)=1+\kappa(\phi,\psi)(1-t)^{1-\phi}+\ldots,\quad\quad 
\mu(t)=
\frac{1}{2\kappa(\phi,\psi)}\left(\theta_1+\theta_t-\phi\right)
(1-t)^{-1+\phi}+\ldots , 
\label{eq:asymptoticlambda}
\ee 
where $\kappa(\phi,\psi)$ is a complicated, but known
\eqref{eq:painleveasymp}, function of the monodromies, and the ellipses
denote subdominant terms in the assumption $0<\Re\phi<1$. Fixed this, the
action is now a function of the condition at the upper limit of
integration, which is the position of the apparent singularity
$\lambda$ and the real singularity at $t$. 

Now, the upper limit of the integration can also be fixed by the
application we have in mind. The Heun equation has four singular regular
points, whereas the Garnier system has five regular points, where one of
them, at $\lambda$ is an apparent singularity. We can then envision a
condition which the apparent singularity coincides with one of the other
singular points, say $t$, and the value of $\theta_t$ is shifted by one.
The condition $\lambda(t_0)=t_0$ seems then natural, and to fix the
values for $K(t_0)$ and $\mu(t_0)$ we consider the limit $\lambda\goesto
t$ in \eqref{eq:heun_canonical}. Then we have $K(t_0)-\mu(t_0)=K_0$
and by taking the same limit in the Hamiltonian \eqref{eq:painleve6h}, we
find: 
\be
\lambda(t_0)=t_0,\quad\quad \mu(t_0)=-\frac{K_0}{\theta_t-1}.
\ee
One can then study the asymptotics of the Painlevé system with these
initial conditions and extract the values of $\phi$ and $\psi$ in the
asymptotic limit $t\goesto 1$. We will leave the full
numerical investigation to future work.

Instead, let us remind that the action $S$ in
\eqref{eq:classical_painleve} is a solution of the Hamilton-Jacobi
equation, and so it implements the canonical transformation between
$\lambda$ and $\mu$ and the monodromy parameters $\phi$ and $\psi$. On
the solutions of the equations of motion, the variations of the action
only depend on the initial and final points of the
trajectory. So,
\be
\psi = -\frac{\partial S}{\partial \phi}=-\left.\mu
\frac{\partial \lambda}{\partial\phi}\right|_{1}^{t_0},
\label{eq:monodromyphipsi}
\ee
where the $t\goesto 1$ limit has to be taken with care. With the
asymptotic conditions \eqref{eq:asymptoticlambda}, the integral has the
logarithmic divergence for $t\goesto 1$: 
\be
S=\frac{1}{4}((\theta_1+\theta_t)^2-\phi^2)\log(1-t)+\ldots
\ee
and this term will be subtracted from the form of the action in order for
us to obtain a finite result for \eqref{eq:monodromyphipsi}. One
should point out that the ellipsis is not analytic at $t=1$, but
still vanishes in the limit. Since the term subtracted is a function of
$t$ alone, the regularized action will still be minimized by the solutions
of the Painlevé and still solve the Hamilton-Jacobi equations. Similar
comments were made in \cite{Litvinov:2013sxa}, although the end result
above is better suited for the isomonodromy problem of the Garnier system.

The specifics of the system renders the usual tools used to study the
isomonodromy problem less than perfect. For instance, one has the
definition of the $\tau$-function for the Painlevé flow:
\be
\frac{d}{dt}\log\tau(t)=K(\lambda(t),\mu(t),t),
\ee
where we use the convention in \cite{Iwasaki:1991}. Other definitions,
such as found in
\cite{Jimbo:1981-2,Jimbo:1982,Hitchin:1997zz,Gamayun:2012ma} differ by an
explicit function of time. From this definition one has the immediate
interpretation of the $\tau$ function as (exponential of) the classical
action for zero momentum configurations. These solutions are of
importance to the theory of uniformization and to special solutions of
Painlevé VI (see \cite{Gamayun:2012ma}). For the case at hand, however,
one would like to introduce a ``generalized'' $\tau$-function, integrating
the whole Lagrangean. 

\begin{figure}[hbt]
\begin{center}
 \mbox{\includegraphics[width=0.7\textwidth]{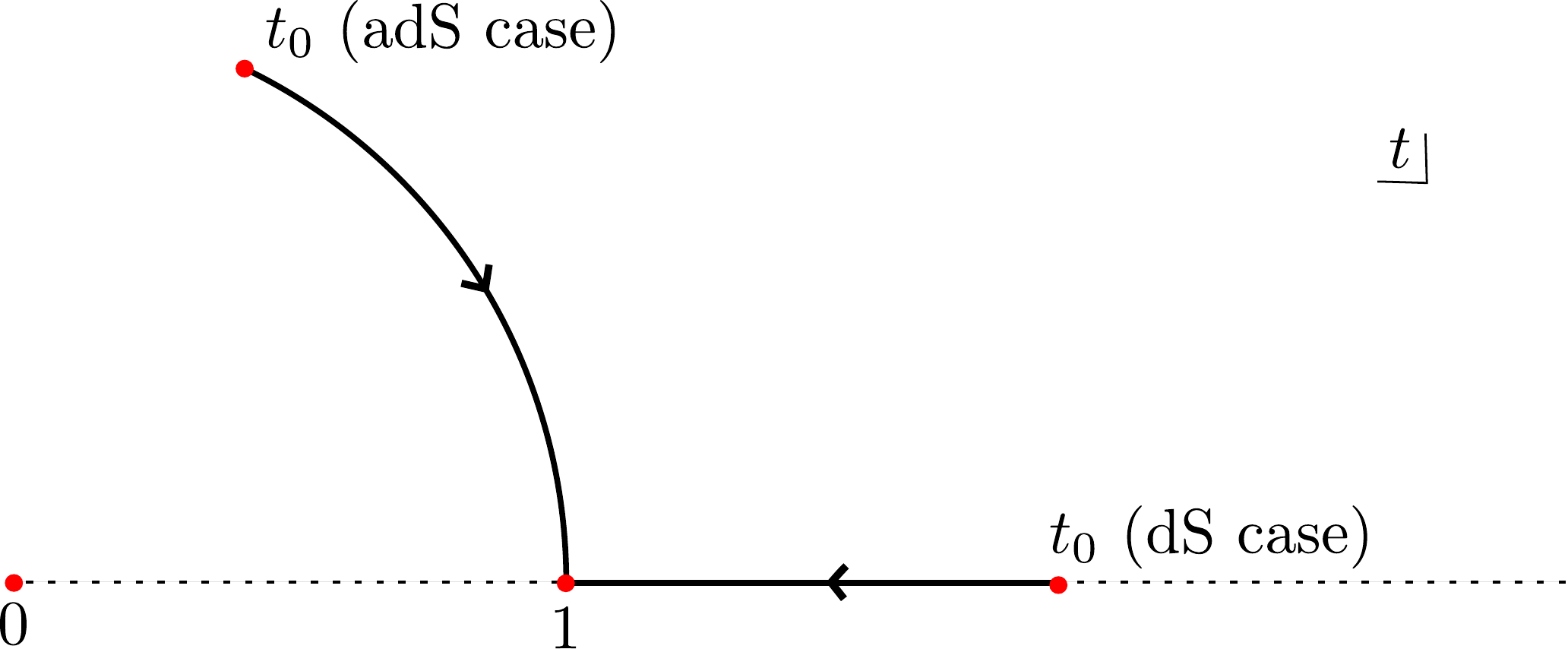}}
\end{center}
\caption{The ``real'' curves in the complex $t$ plane where the
  parameters have a physical interpretation. It is an arc of the unit
  circle for anti-de Sitte and a line segment for the de Sitter case.}
\label{fig:t-plane}
\end{figure}

Finally, one sees from this theory a strange symmetry of the scattering
process. Because the flow of $t$ is defined on the complex plane, we can
always restrict the movement to the ``real'' submanifold where $t$ has a
physical interpretation. We use the word ``real'' loosely here: for the
anti-de Sitter case, we recall that $t_0$ is a phase. One can then
consider a line of ``real phases'' linking $t_0$ to $1$, as in the Figure
\ref{fig:t-plane}. For each point of, say, the arc linking $t_0$ to $1$,
one has a pair of parameters $(t,K(t))$ with the same scattering
properties as $(t_0,K_0)$. The values for $\lambda(t),\mu(t)$ won't
matter for the scattering because the singularity at $\lambda$ is
apparent. The de Sitter case is much the same, but now the line of
physical parameters is also real. That this symmetry is deeply linked to
the Painlevé transcendent is something of a surprise and surely
its understanding deserves more work.

\section{The Generic Scattering}

In essence, the procedure outlined in the preceeding sections does
give scattering elements in terms of the changing basis matrices
$g_i^{-1}g_j$. Apart from computing transmission and reflecting
coefficients to black hole solutions, the full set of monodromies can
be used to compute scattering elements between different asymptotic
regions.

\begin{figure}[htb] 
\begin{center}
\mbox{\includegraphics[width=0.45\textwidth]{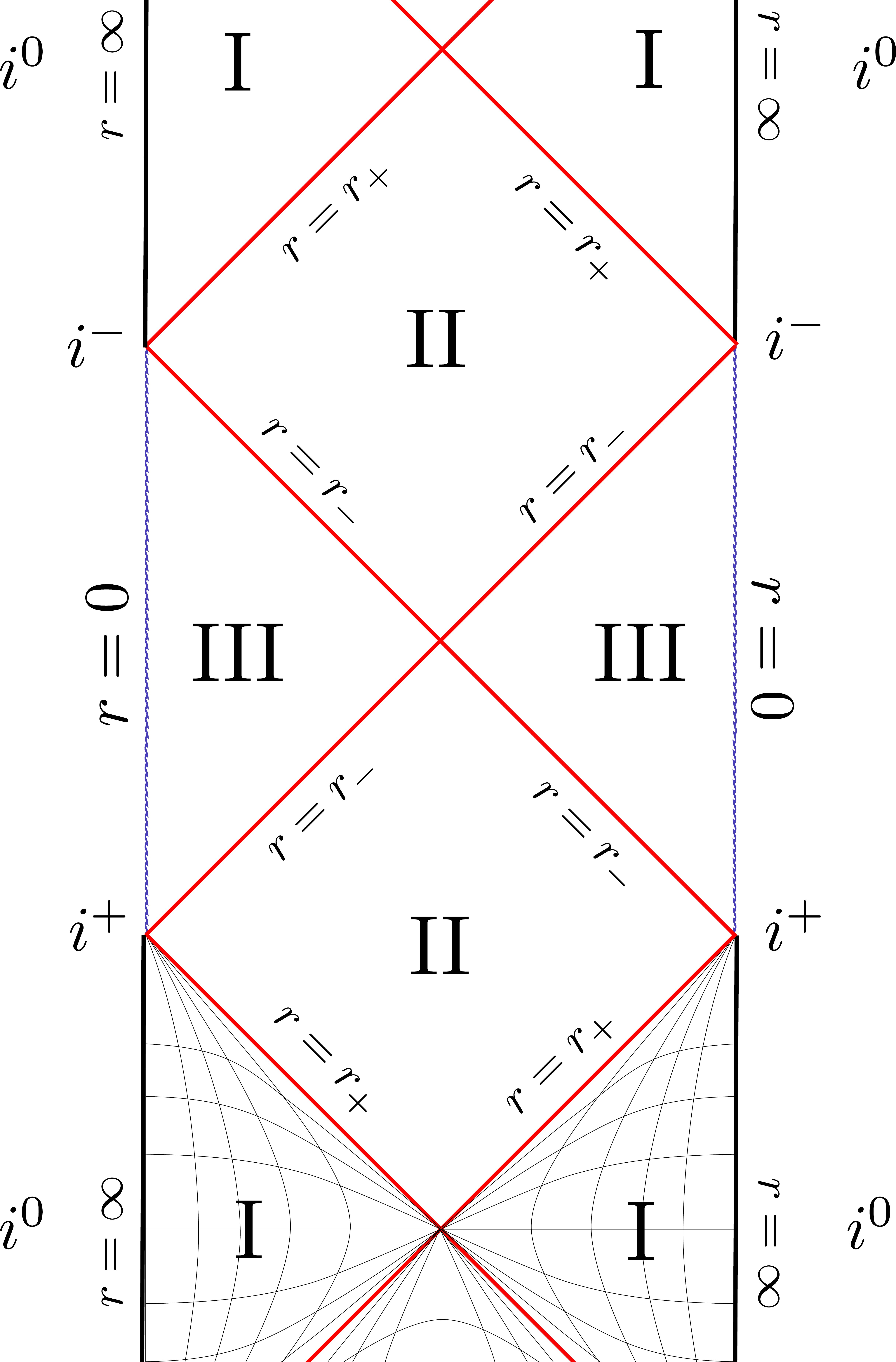}}
\end{center}
\caption{The causal (Penrose) diagram of the (eternal) Kerr black
hole in AdS space. To each asymptotic region one assigns a Hilbert space
${\cal H}$.}
\label{fig:kerr-ads}
\end{figure}

\begin{figure}[htb]
 \begin{center}
  \mbox{\includegraphics[width=\textwidth]{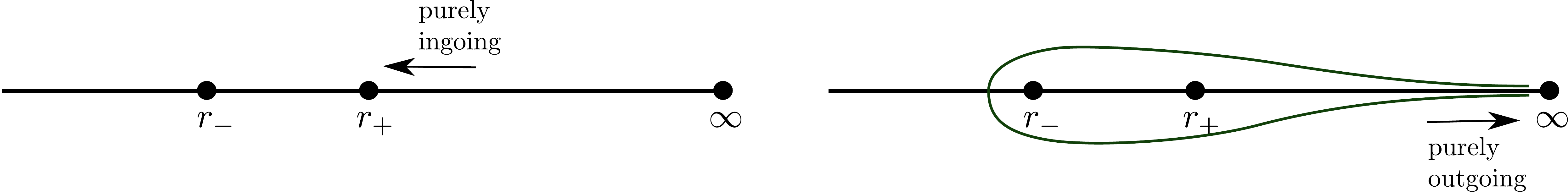}}
 \end{center}
\caption{The schematics of scattering. In the left hand side, the
constrain that the solution is ``purely ingoing'' at $r=r_+$. In the
right-hand side, the monodromy associated with a solution that emerges at
a different region $I$.}
\label{fig:scatt_schem}
\end{figure}

We will be interested in the Kerr-AdS case where the real singularity
points correspond to the horizons and spatial infinity, and all singular
points are regular. Usually, one is
interested in the region $r_+<r<\infty$, where ``classical'' movement
takes
place. The interpretation of elements of $g_i^{-1}g_j$ as scattering
elements \eqref{eq:scattering_matrix} comes about because one chooses the
purely ``ingoing'' solution 
at the singular point $r_+$, as in Figure \ref{fig:scatt_schem}. One then
associates with the matrix an
oriented path, or a graph, between the two singularities. The transmission
and reflection coefficients should be seen as a linear map between the
asymptotic region and itself, the S-matrix.

However, the black hole metric in general can be analytically
continued past $r_+$. In terms of generic Kruskal coordinates, which
follow geodesics, the region $r>r_+$ is but one of many different
asymptotically adS regions. This is schematically presented through
the Penrose diagram in Figure \ref{fig:kerr-ads}. The metric for the other
regions is obtained by analytic continuation: each corresponds to a
different leaf of the Riemann surface determined by the solution of the
Einstein equations.

In other words, the metric in the other asymptotic regions is obtained
by analytic continuation ``around'' the singularity. In the Kerr-AdS
case, as in many others, the form of the metric in the ``inside'' of the
black hole, or, more appropriately, the regions II and III in Figure
\ref{fig:kerr-ads}, is obtained by selecting $r_-<r<r_+$ in the form of
the metric. The passage between different coordinate patches are
constructed from global, Kruskal-like coordinates, but the end result can
be understood as analytical continuation around the singular points
$r=r_+$ and $r=r_-$. Provided one keeps to the real line, the metric thus
obtained will also be real and describe the other regions of the black
hole. Generically, as we go around the complex $z$ plane back to the
asymptotic region of large $z$, the solution of the Klein-Gordon equation
will pick a monodromy. But, by the construction outlined, the large $z$
region will belong to some other asymptotically adS region. This is
illustrated in the right hand side of Figure \ref{fig:scatt_schem}. 

How to interpret the monodromy matrix then? If one chooses a basis like
pictured in Figure \ref{fig:scatt_schem}, where one has a ``purely
outgoing'' wave at spatial infinity, the monodromy path is naturally
associated with the wave travelling along regions II and III in
\ref{fig:kerr-ads}. Since the monodromy path cannot be disentagled, one
should return to a different leaf of the Riemann surface parametrized by
complex $r$. Physically, one returns to a different asymptotic region
I. The monodromy matrix then allows us to compute the scattering
coefficients between different asymptotic regions. As a result, one
should assign to each asymptotic region I a Hilbert space ${\cal H}_i$.
Theoretically, one could also assign a Hilbert space to the singularity
region $r=0$, but it is less clear what is that the scattering
coefficients are measuring, and we will omit this discussion in the
following.

A similar construct was outlined before
\cite{CarneirodaCunha:2001jf,Maldacena:2001kr} in other to describe
local states in the interior of the black hole. There, as in here, one
points out the necessity of the other asymptotic regions in order to
describe the interior of the black hole.
In \cite{Fidkowski:2003nf},
this type of scattering was used to probe the region near the
singularity. In all of those discussions, the number of regions is
limited to two, because one is dealing with
Schwarzschild-type of black holes. The causal diagram for black holes
with zero angular momentum is singular in which there are only two
separate asymptotic regions, as in the Kerr black-hole. Taking generic
lessons from the analysis of these singular cases may be dangerous.

The method outlined here should work for the generic case. In the case
where one obtains a hypergeometric equation for the radial part,
incidentally, corresponds to cases of extreme black holes where the
extra asymptotic regions are also missing. This can be further
verified by studing the extremal limit where $t\goesto 1$ in
\eqref{eq:heun_canonical}. Dealing with the Heun
equation is paramount to study the phenomenon of scattering {\it
  through} the black hole. We hope to address this quantitatively in the
future.

There is, however, some generic conclusions one can take from the
monodromy method without resorting to numerical computations on the
Painlevé transcendents. First and foremost, the isomonodry points to a
hidden symmetry of the Klein-Gordon equation, where different
accessory parameters give off the same scattering elements. Presumably
this can be better understood from twistor methods. 

Secondly, the subgroup of ${\rm SL(2,\mathbb{C})}$ generated by $M_i$
will in general have no closed orbits. A Kleinian group \cite{Bers:1974}
is defined as a discrete subgroup of ${\rm SL(2,\mathbb{C})}$.
Classically, Kleinian groups appear as monodromies of algebraic solutions
of the Heun equation, which, in turn, are related to crystallographic
groups. One famous family of examples are the triangle groups of
tesselations of the Poincaré disk. These appear in the special case where
the conjugacy classes of the monodromy matrices are rational numbers (and
purely imaginary). 

When we are in the generic case where this doesn't happen, but still
the traces $m_i$ are purely imaginary, the orbits of the group
generated by the monodromies describe dense circles in the
group. Generically, then, the amplitudes of processes interpolating
between different will interfere destructively and cancel out. Because
of this, any process that has to be traced over an infinite number of
asymptotic regions will {\it appear to be} unitary. Therefore, in order
to measure the effects that those different asymptotic regions have in
the scattering process, one has to go beyond quantum mechanics and study
field theoretic (higher order) correlations. This sort of ``black hole
complementarity'' stems from the attribution that different asymptotic
regions should have independent Hilbert spaces. The question of locality
in quantum mechanics is sufficiently muddled in order for this not to be a
trivial assumption. At any rate, one sees no reason in order that
effects from these regions should not play a significant role in field
theoretical processes, where one goes beyond the two-point function
evolution. 

\section{Discussion}

In this paper we discussed the application of the isomonodromy method to
the calculation of scattering amplitudes of a generic Kerr-NUT-(A)dS
black hole. These spacetimes have the necessary algebraic properties to
ensure separability of the wave equation, which can be cast as the
problem of two (coupled) ordinary differential equations of the Fuchsian
type. The algebraic property turns out to be a Killing-Yano tensor, which
is closely tied to Petrov type D spacetimes, and are tied to a
subjacent twistorial structure. The fact that we had, for generic values
for the curvature coupling $\xi$, mass $M$, angular momentum $a=J/M$, NUT
charge $b$ and cosmological constant $\Lambda$, a Fuchsian equation is a
remarkable fact in itself. We found further that, while the generic
coupling has 5 regular singular points for the radial
equation\footnote{\footnotesize This equation has been associated with the
name of Böcher in classical treatises \cite{Forsyth:1902}, which states
that all equations of classical mathematical physics can be derived from
it through the process of confluence. It is amusing that the same seems to
hold for black hole scattering!}, the conformally coupled case $\xi=1/6$
has one apparent singularity, and the ensuing equation is of the Heun type
\cite{Batic2007}.

Since the connection problem for the Heun equation is an open
problem, we turned to the isomonodromy method. Although the method has
historical ties to Fuchsian equations, it is more suited to the type of
Schlesinger system discussed in section 3. The difference is that the
latter has apparent singularities with trivial monodromies, and it is
the dynamics of the apparent singularities which has the Painlevé
property and hence can be ``integrated'' in the sense that the ensuing
differential equations define uniquely a function on the complex plane.
The case of four singular points, the Heun case, results in the Painlevé
VI equation for the dynamic of the position of the apparent singularity.
Other cases (and other Painlevé equations) are obtained from confluence,
and this is a very interesting problem in itself, deeply tied with the
so-called Stokes phenomenon with applications in the scattering of Kerr
black holes in flat spaces and the representation of the Virasoro algebra
for Liouville field theory. This should be the object of future studies.

The application of the isomonodromy method to the connection problem of
the Heun equation is somewhat simpler than the analogous problem of
monodromy of the four regular singular point Schlesinger system. The
guidelines of solving it where outlined in \cite{Litvinov:2013sxa},
using the Hamilton-Jacobi method described in section 4 and 5. Since 
the application for black hole scattering selects naturally an initial
condition for the Painlevé system, the action needs only regularization
at one particular point, which we chose to be $t=1$. There is a
mathematical relation between the action and the $\tau$ function
introduced in field theoretic applications \cite{Jimbo1981b}, but for
the generic black hole scattering it is an extension of the latter. The
result can be obtained numerically and will be presented separately.

In section 6 we present a discussion of how the knowledge of the full
monodromy problem can shed light on aspects of black hole
complementarity, specially questions of unitarity and scattering between
different asymptotic regions. Mathematically there seems to be a deep
connection to the theta function associated to an isomonodromy flow
\cite{Jimbo:1981-2}, which can be used to detect deviations from the
purely unitary scattering. This feature is particular to the Heun
equation, and do not show in cases where it reduces to the hypergeometric
case. Amusingly, the hypergeometric cases are obtained when the spacetime
does not display multiple asymptotic regions, like in the extremal black
hole and $AdS_2\times S^2$ cases. The study of the generic case of
scattering will surely be of impact not only to black hole physics, but
also to generic correlations in dual systems described in the
gauge/gravity correspondence.

\section*{Acknowledgements}

The authors would like to thank Amílcar de Queiroz, Dmitry Melnikov, Mark
Mineev-Weinstein, A. Yu. Morozov and Marc Casals for useful
discussions and comments. Fábio Novaes acknowledges partial support
from CNPq. 

\section*{Appendix: Asymptotics of the Schlesinger system and Painlevé VI}

Here we list the relevant results in the asymptotics of Painlevé VI as
studied by \cite{Jimbo:1982} and \cite{Boalch:2005} and listed in
\cite{Guzzetti2011}. The problem was also considered in
\cite{Menotti:2014kra}. In order to study the monodromy near the point
$t=0$, consider the Schlesinger equations for the Heun system
\eqref{eq:schlesinger_heun}. In the approximation we have
\be
\frac{dA_0}{dt}=\frac{[A_t,A_0]}{t},\quad
\frac{dA_1}{dt}=-[A_t,A_1],\quad
\frac{dA_t}{dt}\approx \frac{[A_0,A_t]}{t}+{\cal O}(t^0).
\ee
This means that, near $t=0$, $A_t$ and $A_0$ have a logarithmic
divergence:
\be
A_0\approx t^{\Lambda}A^0_0 t^{-\Lambda},\quad\text{ and }\quad
A_t\approx t^{\Lambda}A^0_t t^{-\Lambda}\quad,\text{ where }\Lambda =
A^0_0+A^0_t, 
\ee
whereas $A_1$ has a continuous limit as $t\goesto 0$. In terms of the
fundamental matrix $\Phi(z,t)$ in \eqref{eq:puregauge},
the system splits into an equation for $\Phi_0(z)=\lim_{t\goesto
0}\Phi(z,t)$ and another for $\Phi_1(z)=\lim_{t\goesto 0}t^{-\Lambda}
\Phi(tz,t)$:
\be
\frac{d\Phi_0}{dz}=\left(\frac{A_1^0}{z-1}+
\frac{\Lambda}{z}\right)\Phi_0,\quad\quad
\frac{d\Phi_1}{dz}=\left(\frac{A_0^0}{z}+\frac{A_t^1}{z-1}\right)\Phi_1.
\ee
Each problem give a hypergeometric connection. Assuming the general case
where there's no integer difference between the exponents, the solutions
are:
\begin{gather}
\Phi_0=\Phi(\frac{1}{2}(\theta_\infty-\theta_1-\phi);
-\frac{1}{2}(\theta_\infty+\theta_1+\phi);1-\phi;z)z^{-\phi/2}(z-1)^{
-\theta_1/2} \\
\Phi_1=G_1\Phi(-\frac{1}{2}(\theta_0+\theta_t+\phi);
-\frac{1}{2}(\theta_0+\theta_t-\phi);1-\theta_0;z)C_1z^{-\theta_0/2}
(z-1)^ {-\theta_t/2}.
\end{gather}
With the hypergeometric fundamental solution given by:
\be
\Phi(\alpha,\beta;\gamma;z)=
\begin{pmatrix}
 \Phi_{11} & \Phi_{12} \\
 \Phi_{21} & \Phi_{22}
\end{pmatrix}
z^{-\left(\begin{smallmatrix}
     \alpha & 0 \\
      0 & \beta
    \end{smallmatrix}\right)}.
\ee
and
\be
\begin{gathered}
\Phi_{11}={_2F_1}(\alpha,\alpha-\gamma+1;\alpha-\beta;\frac{1}{z}) \\
\Phi_{12}= \frac{\beta(\beta-\gamma+1)}{(\beta-\alpha)(\beta-\alpha+1)}
\frac{1}{z}
{_2F_1}(\beta+1,\beta-\gamma+2;\beta-\alpha+2;\frac{1}{z}) \\
\Phi_{21}=\frac{\alpha(\alpha-\gamma+1)}{(\alpha-\beta)(\alpha-\beta+1)}
\frac { 1 } { z }
_2F_1(\alpha+1,\alpha-\gamma+2;\alpha-\beta+2;\frac{1}{z}) \\
\Phi_{22}={_2F_1}(\beta,\beta-\gamma+1;\beta-\alpha;\frac{1}{z}).
\end{gathered}
\ee
The constants are given by:
\be
\alpha=\frac{1}{2}(\theta_\infty-\theta_1-\phi),\quad
\beta=\frac{1}{2}(-\theta_\infty-\theta_1-\phi),\quad
\gamma=1-\phi.
\ee
The asymptotic of the hypergeometrics are as:
\be
 Y(\alpha,\beta,\gamma;z) =
\begin{cases}
 G^{(0)}_{\alpha\beta\gamma}
(1+{\cal O}(z))z^{
\left(\begin{smallmatrix}
 1-\gamma & 0 \\
 0 & 0
\end{smallmatrix}\right)
}C^{(0)}_{\alpha\beta\gamma}, & z\goesto 0; \\
G^{(1)}_{\alpha\beta\gamma}
(1+{\cal O}(z-1))(z-1)^{\left(
\begin{smallmatrix}
 \gamma-\alpha-\beta-1 & 0 \\
 0 & 0
\end{smallmatrix}\right)
}C^{(1)}_{\alpha\beta\gamma}, & z\goesto 1; \\
(1+{\cal O}(z^{-1}))z^{\left(
\begin{smallmatrix}
 -\alpha & 0 \\
 0 & -\beta
\end{smallmatrix}\right)
}, & z\goesto \infty. \\
\end{cases}
\ee
The matrices 
\be
G^{(0)}_{\alpha\beta\gamma}=
\frac{1}{\beta-\alpha}
\begin{pmatrix}
 \beta-\gamma+1 & \beta \\
 \alpha-\gamma+1 & \alpha
\end{pmatrix},
\quad\quad
G^{(1)}_{\alpha\beta\gamma}=
\frac{1}{\beta-\alpha}
\begin{pmatrix}
 1 & \beta(\beta-\gamma) \\
 1 & \alpha(\alpha-\gamma)
\end{pmatrix}.
\ee
And the connection matrices:
\be
\begin{gathered}
 C^{(0)}_{\alpha\beta\gamma}=
\begin{pmatrix}
 e^{-\pi
i(\alpha-\gamma+1)}\frac{\Gamma(\gamma-1)\Gamma(\alpha-\beta+1)}{
\Gamma(\gamma-\beta)\Gamma(\alpha)} & 
e^{-\pi
i(\beta-\gamma+1)}\frac{\Gamma(\gamma-1)\Gamma(\beta-\alpha+1)}{
\Gamma(\gamma-\alpha)\Gamma(\beta)} \\
e^{-\pi
i\alpha}\frac{\Gamma(1-\gamma)\Gamma(\alpha-\beta+1)}{
\Gamma(1-\beta)\Gamma(\alpha-\gamma+1)} &
-e^{-\pi
i\beta}\frac{\Gamma(1-\gamma)\Gamma(\beta-\alpha+1)}{
\Gamma(1-\alpha)\Gamma(\beta-\gamma+1)}
\end{pmatrix}, \\
C^{(1)}_{\alpha\beta\gamma}=
\begin{pmatrix}
 -\frac{\Gamma(\alpha+\beta-\gamma+1)\Gamma(\alpha-\beta+1)}{
\Gamma(\alpha-\gamma+1)\Gamma(\alpha)} & 
\frac{\Gamma(\alpha+\beta-\gamma+1)\Gamma(\beta-\alpha+1)}{
\Gamma(\beta-\gamma+1)\Gamma(\beta)} \\
-e^{-\pi
i(\gamma-\alpha-\beta-1)}\frac{
\Gamma(\gamma-\alpha-\beta-1)\Gamma(\alpha-\beta+1) } {
\Gamma(1-\beta)\Gamma(\gamma-\beta)} &
e^{-\pi
i(\gamma-\alpha-\beta-1)}\frac{
\Gamma(\gamma-\alpha-\beta-1)\Gamma(\beta-\alpha+1) } {
\Gamma(1-\alpha)\Gamma(\gamma-\alpha)}
\end{pmatrix}.
\end{gathered}
\ee

The asymptotics of the $A^t$ are worked out it \cite{Jimbo:1982}:
\begin{gather}
\Lambda +\frac{1}{2}\phi\mathbb{I}\simeq \frac{1}{4\theta_\infty}
\begin{pmatrix}
 (-\theta_\infty-\infty_1+\phi) (\theta_\infty-\infty_1-\phi) &
 (-\theta_\infty-\infty_1+\phi) (\theta_\infty+\infty_1+\phi) \\
 (\theta_\infty-\infty_1+\phi) (\theta_\infty-\infty_1-\phi) &
 (\theta_\infty-\infty_1+\phi) (\theta_\infty+\infty_1+\phi)
\end{pmatrix};
\\
A_1^0 +\frac{1}{2}\theta_1\mathbb{I}\simeq \frac{1}{4\theta_\infty}
\begin{pmatrix}
 -(\theta_\infty-\theta_1)^2+\phi^2 & 
 (\theta_\infty+\theta_1)^2-\phi^2 \\
-(\theta_\infty-\theta_1)^2+\phi^2 & 
 (\theta_\infty+\theta_1)^2-\phi^2 \\
\end{pmatrix};
\\
A_0^0+\frac{1}{2}\theta_0\mathbb{I}=G_1 \frac{1}{4\phi}
\begin{pmatrix}
 (\theta_0-\theta_t+\phi)(\theta_0+\theta_t+\phi) & 
 (\theta_0-\theta_t+\phi)(-\theta_0-\theta_t+\phi) \\
 (\theta_0-\theta_t-\phi)(\theta_0+\theta_t+\phi) &
 (\theta_0-\theta_t-\phi)(-\theta_0-\theta_t+\phi) 
\end{pmatrix}
G_1^{-1};
\\
A_t^0+\frac{1}{2}\theta_t\mathbb{I}=G_1\frac{1}{4\phi}
\begin{pmatrix}
 (\theta_t+\phi)^2-\theta_0 & -(\theta_t-\phi)^2+\theta_0^2 \\
 (\theta_t+\phi)^2-\theta_0 & -(\theta_t-\phi)^2+\theta_0^2
\end{pmatrix}
G_1^{-1}.
\end{gather}
The matrix is given by
\be
G_1=G^{(0)}_{\alpha\beta\gamma}
\begin{pmatrix}
 1 & 0 \\
 0 & -\hat{s}^{-1}
\end{pmatrix}, 
\ee
\be
\begin{gathered}
\hat{s}=\frac{\Gamma(1-\phi)^2\Gamma({\textstyle \frac{1}{2}}(\theta_0
+\theta_t+\phi)+1)\Gamma({\textstyle \frac{1}{2}}(-\theta_0
+\theta_t+\phi)+1)}{\Gamma(1+\phi)^2\Gamma({\textstyle
\frac{1}{2}}(\theta_0
+\theta_t-\phi)+1)\Gamma({\textstyle \frac{1}{2}}(-\theta_0
+\theta_t-\phi)+1)}\times \\
\frac{\Gamma({\textstyle \frac{1}{2}}(\theta_\infty
+\theta_1+\phi)+1)\Gamma({\textstyle \frac{1}{2}}(-\theta_\infty
+\theta_1+\phi)+1)}{\Gamma({\textstyle
\frac{1}{2}}(\theta_\infty
+\theta_1-\phi)+1)\Gamma({\textstyle \frac{1}{2}}(-\theta_\infty
+\theta_1-\phi)+1)}s,
\end{gathered}
\ee
with $s$ parameter given by
\be
\begin{gathered}
4\sin\frac{\pi}{2}
(\theta_0+\theta_t\mp\phi)\sin\frac{\pi}{2}(\theta_0-\infty_t\pm\phi)
\sin\frac{\pi}{2}(\theta_\infty+\theta_1\mp\phi)
\sin\frac{\pi}{2}(\theta_\infty-\theta_1\pm\phi)
s^\pm= \\
(\pm
i\sin\pi\phi\cos\pi\sigma_{1t}-\cos\pi\theta_t\cos\pi\theta_\infty
-\cos\pi\theta_0\cos\pi\theta_1)e^{\pm\pi i\phi}
\\
\pm i\sin\pi\phi\cos\pi\sigma_{01}+\cos\pi\theta_t\cos\pi\theta_1+ 
\cos\pi\theta_\infty\cos\pi\theta_0. \\
\end{gathered}
\ee

With these expressions, one can calculate the asymptotic expansion for
the $\tau$ function:
\be
\frac{d}{dt}(t(t-1)\frac{d}{dt}\log\tau(t))=
\theta_\infty (A_t(t))_{22}-\frac{1}{2}\theta_t^2.
\ee
Obtaining \cite{Jimbo:1982}:
\be
\begin{aligned}
\tau(t)\simeq & t^{(\phi^2-\theta_0^2-\theta_t^2)/4}\left[
1+\frac{1}{8\phi^2}(\theta_0^2-\theta_t^2-\phi^2)
(\theta_\infty^2-\theta_1^2 -\phi^2)t- \right. \\ & 
\quad \frac{\hat{s}}{16\phi^2(1+\phi^2)}(\theta_0^2-(\theta_t-\phi)^2)
(\theta_\infty^2-(\theta_1-\phi)^2)t^{1+\phi} \\ & 
\quad\quad \left.
-\frac{\hat{s}^{-1}}{16\phi^2(1-\phi^2)}(\theta_0^2-(\theta_t+\phi)^2)
(\theta_\infty^2-(\theta_1+\phi)^2)t^{1-\phi}
+{\cal O}(|t|^{2(1-\Re\phi)})\right].
\end{aligned}
\ee
For the asymptotics as $t\goesto
1$, one just need to change $\theta_0$ to $\theta_1$, and $\lambda(t)$
to $\lambda(t)-1$. Finally, the asymptotic formula for the Painlevé
transcendent itself, as in \cite{Guzzetti2011}:
\be
\lambda(t)\simeq 1+\frac{(\theta_t-\theta_1+\phi)
(\theta_t+\theta_1+\phi)(\theta_\infty+\theta_0+\phi)}{4
\phi^2(\theta_\infty+\theta_0-\phi)\hat{s}}
(1-t)^{1-\phi}(1+{\cal O}(t^\phi,t^{1-\phi})),
\label{eq:painleveasymp}
\ee
assuming, as always, $0<\Re\phi<1$. 

\providecommand{\href}[2]{#2}\begingroup\raggedright\endgroup

\end{document}